\documentclass[10pt]{article}
\usepackage{amsmath,amsfonts,amssymb,amsthm,bbm}
\usepackage{pdfsync}
\usepackage{graphicx}
\usepackage[latin1]{inputenc}
\usepackage{hyperref}
\usepackage[all]{xy}
\usepackage{stmaryrd}
\usepackage{setspace}

\newcommand{\bit}{\begin{itemize}}
\newcommand{\eit}{\end{itemize}}
\newcommand{\bd}{\begin{description}}
\newcommand{\ed}{\end{description}}

\newcommand{\bc}{\begin{center}}
\newcommand{\ec}{\end{center}}
\newcommand{\Ref}[1]{(\ref{#1})}

\newcommand{\C}{{\mathbb C}}

\newcommand{\R}{{\mathbb R}}

\newcommand{\hh}{{\cal H}}

\newcommand{\be}{\begin{equation}}
\newcommand{\ee}{\end{equation}}
\newcommand{\bea}{\begin{eqnarray}}
\newcommand{\eea}{\end{eqnarray}}
\newcommand{\bs}{\begin{subequations}}
\newcommand{\es}{\end{subequations}}
\newcommand{\nn}{\nonumber}

\newcommand{\f}{\frac}
\newcommand{\tl}{\tilde}
\def\p{\partial}

\newcommand{\re}{\mathrm{Re}}
\newcommand{\im}{\mathrm{Im}}

\newcommand{\na}{\nabla}

\newcommand{\mat}[4]{\left( \begin{array}{cc}{#1}&{#2}\\{#3}&{#4}\end{array} \right)}

\renewcommand{\a}{\alpha} \renewcommand{\b}{\beta} \newcommand{\g}{\gamma}
\renewcommand{\d}{\delta}  \newcommand{\eps}{\epsilon}  \newcommand{\z}{\zeta}
 \renewcommand{\th}{\theta}    \renewcommand{\k}{\kappa}  \renewcommand{\l}{\lambda}
\let\m=\mu  \let\n=\nu  \let\r=\rho \newcommand{\s}{\sigma}  \renewcommand{\t}{\tau}   \let\vphi=\varphi  \let\om=\omega
 \let\D=\Delta  \let\Th=\Theta \let\L=\Lambda \let\Si=\Sigma

\def\cN{N}

\newcommand{\tE}{\lefteqn{\smash{\mathop{\vphantom{<}}\limits^{\;\sim}}}E}
\newcommand{\tP}{\lefteqn{\smash{\mathop{\vphantom{<}}\limits^{\;\sim}}}P}

\newcommand{\Et}{\lefteqn{\smash{\mathop{\vphantom{\Bigl(}}\limits_{\sim}
\atop \ }}E}

\newcommand{\tNn}{\lefteqn{\smash{\mathop{\vphantom{\Bigl(}}\limits_{\,\sim}
\atop}}{\, N}}

\newcommand{\bB}{\mathbb{B}}
\newcommand{\bG}{\mathbb{G}}
\newcommand{\scri}{\cal I}

\newcommand{\os}[1]{\overset{\circ}{#1}}

\begin{document}

\title{\bf Sachs' free data in real connection variables}

\author{\Large{Elena De Paoli$^1$ and Simone Speziale$^2$}
\smallskip \\ \small{$^1$Dip. di Fisica, Univ. di Roma 3, Via della Vasca Navale 84, 00146 Roma, Italy, and}  \\
 \small{Dip. di Fisica, Univ. di Roma La Sapienza, Piazzale A. Moro 2, 00185 Roma, Italy}  \\
\small{$^2$ Aix Marseille Univ., Univ. de Toulon, CNRS, CPT, UMR 7332, 13288 Marseille, France}
}
\date{\today}

\maketitle

\begin{abstract}
\noindent 
We discuss the Hamiltonian dynamics of general relativity with real connection variables on a null foliation, and use the Newman-Penrose formalism to shed light on the geometric meaning of the various constraints. We identify the equivalent of Sachs'
constraint-free initial data as projections of connection components related to null rotations, i.e. the translational part of the ISO(2) group stabilising the internal null direction soldered to the hypersurface. 
A pair of second-class constraints reduces these connection components to the shear of a null geodesic congruence, thus establishing equivalence with the second-order formalism, which we show in details at the level of symplectic potentials. A special feature of the first-order formulation is that Sachs' propagating equations for the shear, away from the initial hypersurface, are turned into tertiary constraints; their role is to preserve the relation between connection and shear under retarded time evolution. The conversion of wave-like propagating equations into constraints is possible thanks to an algebraic Bianchi identity; the same one that allows one to describe the radiative data at future null infinity in terms of a shear of a (non-geodesic) asymptotic null vector field in the physical spacetime. 
Finally, we compute the modification to the spin coefficients and the null congruence in the presence of torsion.
\end{abstract}

\tableofcontents

\section{Introduction}
Null foliations play an important role in general relativity.
Among their special features, they admit a gauge-fixing for which the Einstein's equations can be integrated hierarchically, and constraint-free initial data identified, as shown by Sachs \cite{Sachs62}; and provide a framework for the description of gravitational radiation from isolated systems and of conserved charges, starting from the seminal work of Sachs, of Bondi, van der Burg and Metzner (henceforth BMS), Newman and Penrose (NP), Geroch and Ashtekar \cite{Bondi:1960jsa,Sachs:1962wk, BMS,NP62,Penrose:1980yx,Penrose:1962ij,Penrose:1965am,Geroch:1977jn, Ashtekar:1981hw,Ashtekar:1981bq,Ashtekar:2014zsa} (see also \cite{Winicour16,Wald:1999wa,Barnich:2011mi} and reference therein). These classic results are based on the Einstein-Hilbert action and the spacetime metric as fundamental variable, and provide a clear geometric picture of the physical degrees of freedom of general relativity at the non-linear level. In this paper we wish to understand some of these results 
using a first-order action principle with real connection variables. In particular, we will identify the equivalent of Sachs' free data in terms of some connection components (which will be related to the translational part of the ISO(2) group stabilising the internal null direction soldered to the hypersurface), and highlight some properties of their Hamiltonian dynamics. 

We have several reasons to be interested in this. 
First of all, we know from the work of Ashtekar that the radiative physical degrees of freedom at future null infinity are best described in terms of connections \cite{Ashtekar:1981hw,Ashtekar:1981bq}.\footnote{Another class of null hypersurfaces for which  the connection description plays an important role is the one of isolated horizons \cite{Ashtekar:2004cn,AshtekarQIH,Ghosh:2011fc}.} 
We then wish to provide a connection description of the physical degrees of freedom in the spacetime bulk, 
in the sense of constraint-free initial data for the first-order action.
Secondly, the connection description later led Ashtekar to the famous reformulation of the action principle of general relativity \cite{Ashtekar:1986yd}, which is at the root of loop quantum gravity. This approach to quantising general relativity suggests the use of connections as fundamental fields, instead of the metric. There exists a canonical quantisation scheme that leads to the well-known prediction of quantum discreteness of space \cite{RovelliSmolin94}. This result uses space-like foliations, and the dynamical restriction to the quanta of space imposed by the Hamiltonian constraint are still not explicitly known, none-withstanding constant progress in the field. 
Quantising with analogue connection methods the constraint-free data on null foliations would allow us to study the quantum structure of the physical degrees of freedom directly.\footnote{For recent work towards the same goal but in metric variables, see \cite{Fuchs:2017jyk}.}  
As a preliminary result in this direction, it was shown in \cite{IoNull} that at the kinematical level, discretisations of the 2d space-like metric have quantum area operators with a discrete spectrum given by the helicity quantum numbers. A stronger more recent result appeared in \cite{Wieland:2017cmf}, based on covariant phase space methods and a spinorial boundary term, confirming the discrete area spectrum without a discretisation. What we would like is to extend these results within a Hamiltonian dynamical framework.

The Hamiltonian dynamics of general relativity with real connection variables on a null foliation appeared in \cite{IoSergeyNull},\footnote{For previous studies using complex self-dual connections see e.g. \cite{Goldberg:1992st,dInverno:2006mmy}.} and presents a few intricate structures, like the conversion of what Sachs called the propagating Einstein's equations into (tertiary) constraints. In this paper, we present three results. First, we use the Newman-Penrose formalism to clarify the geometric meaning of the various constraints present in the Hamiltonian structure studied in \cite{IoSergeyNull}. Second, we identify the connection equivalent of Sachs' free data as the `shear-like' components of an affine\footnote{In the sense of being given by an affine connection, a priori non-Levi-Civita, not of being affinely parametrised.} null congruence; we show how they reduce to the shear of a null geodesic congruence in the absence of torsion, and how they are modified in the presence of torsion; we use the Bondi gauge to derive their Dirac brackets, and show the equivalence with the metric formalism at the level of symplectic potentials. Third, we explain the origin and the meaning of the tertiary constraints, and point out that the algebraic Bianchi identity responsible for the conversion of the propagating equations into constraints is the same one that allows the interpretation of the radiative data at future null infinity $\scri^+$ in terms of shear of a (non-geodesic) null vector field `aligned' with $\scri^+$.

The identification of the dynamical part of the connection with null rotations (related on-shell to the shear) is a striking difference with respect to the case of a space-like foliation, because these components form a group, albeit a non-compact one, unlike the dynamical components of the space-like formalism which are boosts (related on-shell to the extrinsic curvature). We have thus two senses in which a null foliation gives a simpler algebra:
the first-class part of the constraint algebra is a genuine Lie algebra (thanks to the fact that the Hamiltonian is second class), and the connection physical degrees of freedom form a group. 

The paper is organised as follows. We first review useful background material on the Hamiltonian structure on null foliations: in Section 2, with metric variables, including the use of Bondi coordinates and identification of constraint-free initial data and their symplectic potential; in Section 3, with real connection variables. In Section 4, we map the non-adapted tetrad used in the Hamiltonian analysis to a doubly-null tetrad, we identify the constraint-free data and study the effect of the constraints on an affine null congruence. We describe the modifications induced by torsion in the case of fermions minimally and non-minimally coupled, as well as for a completely general torsion. We rederive the conversion of the propagating equations into constraints using the Newman-Penrose formalism, and single out one algebraic Bianchi identity responsible for it. In Section 5 we specialise to Bondi coordinates, and discuss the Dirac bracket for the constraint-free data and the equivalence of the symplectic potential with the one in metric variables. We finally highlight that the same algebraic Bianchi identity relevant to the understanding of the tertiary constraints plays an interesting role for radiative data at $\scri^+$. The conclusions in Section 6 contain some perspectives on future work.
We also provide an extensive Appendix with technical material. This includes the detailed relation of our tetrad foliation to the $2+2$ foliation used in the literature, of the metric coefficients we use to those of Sachs and of Newman and Penrose, the explicit expression of all NP spin coefficients in the first-order variables, and some details on the mixing between internal boost gauge-fixing and lapse fixing via radial diffeomorphisms.

For the purposes of this paper, we will mostly restrict attention to local considerations on a single null hypersurface. 
We neglect in particular boundary conditions and surface terms. These carry of course very important physics, and we will come back to them in future research, when we have in mind among other things to explore the phase space and BMS algebra on $\scri^+$ in this formalism.  

We use mostly-plus signature (--+++), with greek letters $\m,\n,\ldots=0,1,2,3$ as spacetime indices, latin indices from the beginning of the alphabet $a,b,\ldots=1,2,3$ as hypersurface indices, capital ones $A,B,\ldots=2,3$ as 2d space-like surface indices; for the internal space, we use $I,J,\ldots = 0,1,2,3$ as internal spacetime indices, $i,j,\ldots=1,2,3$ internal hypersurface indices, $M,N,\ldots=2,3$ internal 2d space-like indices.

\section{Sachs' free data and metric Hamiltonian structure}

Before presenting the first-order connection formulation, let us review some basic facts of the metric formulation, that will be useful in the following:
the details of the Bondi coordinate gauge-fixing, and the description of constraint-free data and their associated symplectic potential.

\begin{figure}[ht]   
  \centering      \includegraphics[width=12cm]{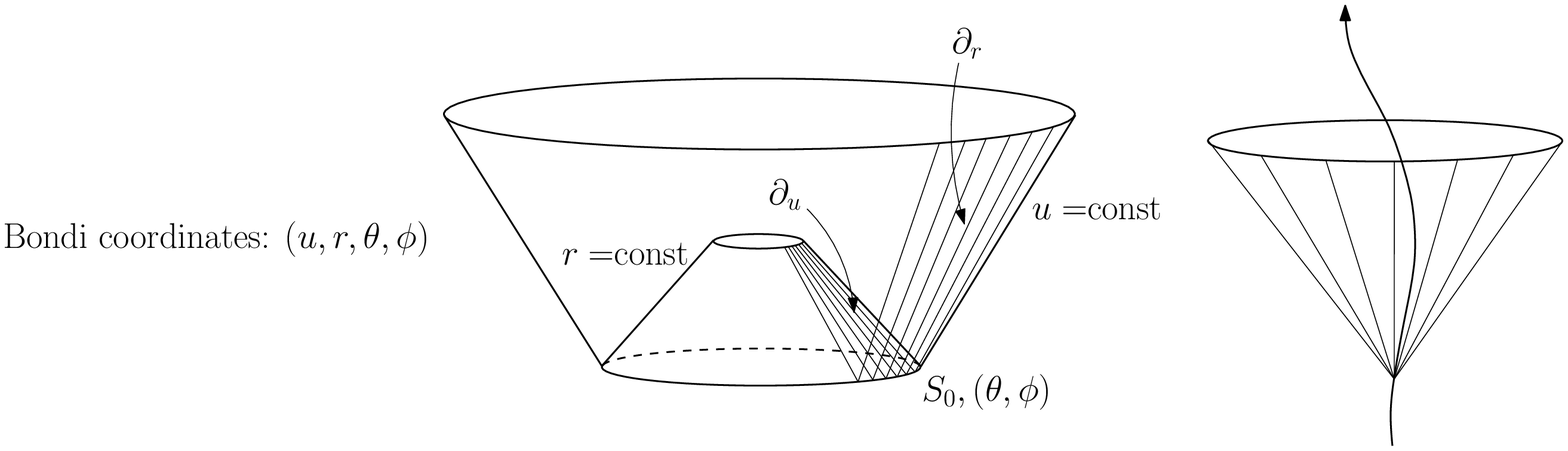}
\caption{\label{Fig} {\small{
Left: \emph{Set-up of the characteristic $2+2$ initial-value problem. Two null hypersurfaces intersect on a space-like 2d surface $S_0$.  When the two null hypersurfaces are intersecting light cones, as in the picture, $S_0$ has topology of a sphere. The (partial) Bondi gauge is such that $(\th,\phi)$ are constant along $\p_r$, and $\p_r$ is null for all values of $u$. On the other hand, $\p_u$ is null at at most one value of $r$, unless the spacetime has special isometries.} Right: \emph{Further requiring suitable regularity conditions one can consider also a local $3+1$ foliation of light-cones generated by a time-like world-line.
} }} }
\end{figure}

The typical set-up is a $2+2$ foliation with a doubly-null initial slice, see Fig. \ref{Fig}. Sachs' constraint-free data for a local evolution can then be identified with the conformal class of the two-dimensional induced metric along the initial slice, or alternatively its shear, plus corner data at the 2d space-like intersection. With some additional regularity assumptions, one can also use a $3+1$ foliation by null cones radiated by a time-like world-line. See \cite{Rendall221,Friedrich1986,FrittelliNull95,ChoquetBruhat:2010ih,Chrusciel:2012xf} for the formal analysis of solutions and existence theorems. Both evolution schemes are typically local because of the development of caustics, however for situations with sufficiently weak gravitational radiation like those of \cite{Christodoulou:1993uv}, null cones can foliate all of spacetime.
A case of special interest is the study of radiating isolated gravitational systems in asymptotically flat spacetimes. In the asymptotic $2+2$ problem, one puts the second null hypersurface at future null infinity $\scri^+$, 
and the foliation describes null hypersurfaces (or null cones) attached to $\scri^+$. In this case the assignment of initial data is subtler (see e.g. \cite{Friedrich:1981at}), because of the compactification involved in the definition of $\scri$. In particular, $\scri^+$ is shear-free by construction. Nonetheless, the data are still described by an asymptotic shear, \emph{transverse} to $\scri^+$ \cite{Penrose:1962ij, NP62,Ashtekar:1981hw}, and Ashtekar's result was to show that these degrees of freedom and the phase space they describe are better thought of in terms of connections living on $\scri^+$, a construction which is useful for the understanding of conserved charges.
Notice that one can not take $\scri^+$ itself as null cone of a $3+1$ foliation, because of the `hole' at $i^+$ where tails and bound states escape null infinity (see e.g. \cite{Geroch:1978us}), nor $\scri^-$ for the same reason.
We will mostly focuse on local properties of null hypersurfaces, and not discuss the non-trivial features associated for instance with boundary data at corners, residual diffeomorphisms, caustics and cone-vertex regularity, for which we refer the reader to literature cited above and below.

%

\subsection{Bondi gauge and Sachs constraint-free initial data}
The Bondi coordinate gauge is 
specified as follows: we take spherical coordinates in a local patch of spacetime, $x^\m = (u,r,\th,\phi)$, with the level sets of $u$ to provide a foliation into null hypersurfaces $\Si$. $du$ is thus a null 1-form, implying the gauge-fixing condition $g^{00} = 0$, and the associated future-pointing null vector $l^\m=-g^{\m\n}\p_\m u$ is tangent to the null geodesics of $\Si$. 
The second gauge condition is to require the angular coordinates $x^A=(\th,\phi)$, $A=2,3$, to be preserved along $r$, i.e. $l^\m \p_\m x^A=0$. This implies $g^{0A}=0$ and makes $r$ a parameter along the null geodesics: the level sets of $r$ thus provide a $2+1$ foliation orthogonal to the null geodesics.
At this point, the metric and its inverse can be conveniently parametrized as follows, 
\be\label{gS}
g_{\m\nu} = \left(\begin{matrix} -e^{2\b}\f Vr + \g_{AB} U^A U^B & -e^{2\b} & -\g_{AB}U^B \\ & 0 & 0 \\ & & \g_{AB} \end{matrix}\right),
\qquad 
g^{\m\nu} = \left(\begin{matrix} 0 & -e^{-2\b} & 0 \\ & e^{-2\b}\f Vr & -e^{-2\b} U^A \\  & & \g^{AB} \end{matrix}\right),
\ee
in terms of seven functions $(\b,V,U^A,\g_{AB})$. Being the coordinates adapted to the $2+2$ foliation defined by $u$ and $r$, $g_{AB}\equiv \g_{AB}$ is the metric induced on the 2d space-like surfaces, and we denote its determinant $\g$ and its inverse $\g^{AB}$. The gauge-fixing has the property that $g^{AB}=\g^{AB}$, so it is analogue to the shift-free (partial) gauge $N^a=0$ for a space-like foliation.
There still remains one coordinate freedom, for which two different choices are customary in the literature: we can require as in \cite{Sachs:1962wk,BMS}  the radial coordinate to be an areal parameter $R$ (called `luminosity distance' by Sachs), namely fix $\sqrt{\g}=R^2 f(\th,\phi)$; or we can follow the Newman-Penrose (NP) literature \cite{NP62, Penrose:1980yx} and require $g^{01}=-1$, with no restrictions on $\g$, 
which makes $r$ an affine parameter for the congruence generated by $l^\m$.
The relation between the two choices is given by ${\p r}/{\p R}=e^{2\b}$. 
As we will review below, $e^{2\b}$ plays the role of the lapse function in the canonical theory, and these two choices correspond to two different gauge-fixings of the radial diffeomorphism constraint. Accordingly, we will denote from now on $e^{2\b}=N>0$.
In the following, we will often keep this last gauge fixing unspecified, for our results to be easily adapted to both choices. We will then refer to the partial gauge-fixing $g^{00}=0=g^{0A}$ as partial Bondi gauge.\footnote{A third option to complete the partial Bondi gauge is to take $dr$ null, so to have also $g^{11}=0$. This choice, used in the original Sachs paper \cite{Sachs62}, is not adapted to the asymptotic problem, and will be not considered in the following.}

To set up the characteristic $2+2$ initial-value problem, one chooses initial data on two null hypersurfaces intersecting on a space-like 2d surface $S_0$, see Fig. \ref{Fig}. Working with a null foliation, any fixed value of $u$ identifies the first null hypersurface. On the other hand, with $r$ affine or areal at most one $r=$constant hypersurface will also be null, for a generic spacetime. Its location can be fixed with a measure-zero gauge-fixing $g^{11}|_{r_0}=0$. 
Then, as shown originally in  \cite{Sachs62} (see also \cite{dInverno:1980kaa,Torre:1985rw,Winicour16}), 
constraint-free initial data for general relativity can be identified with the conformal class of 2d space-like metrics $\g_{AB}$, of which we take 
the uni-modular representative $\check\g_{AB}:=\g^{-1/2}\g_{AB}$; supplemented by boundary data at the corner $S_0$ between the two initial slices.\footnote{Explicitly, Sachs' also fixes the residual hypersurface gauge, and provides the corner data
\be\nn
(\g, \quad \p_u \g, \quad \p_r \g , \quad \p_r U^A)|_{S_0}. 
\ee 
They provide the area of $S_0$, the initial expansion of the null geodesic congruences along the two hypersurfaces, and the non-integrability of the two null directions: $U_A,_1$ gives in these coordinates the Lie bracket among the two normal vectors $\p_u$ and $\p_r$ at $S_0$. 
}
Up to the measure-zero corner data, the two independent components of $\check\g_{AB}$ are the two physical degrees of freedom of general relativity on a null hypersurface. In the associated hierarchical integration scheme, the Hamiltonian constraint can be solved as a radial linear equation for $V$, and one can identify the propagating equations for the constraint-free data as (the traceless part of) the projection of the Einstein's equations on the space-like surface. 
These give the evolution of $\check\g_{AB}$ away from the initial slice.
The price to pay for the identification of constraint-free data is that the dynamical spacetime can be reconstructed only locally in a neighbourhood of the characteristic surface (neighbourhood that may well be smaller then the maximal Cauchy development, see e.g. \cite{Rendall221}),
as caustics develop and stop the validity of the coordinate patch. See e.g. \cite{FrittelliNull95,d2005approaches,MikeNull,Winicour16} for various discussions on this.

The geometric interpretation of the constraint-free data is most commonly given in terms of the shear 
of null geodesic congruences, which is directly determined by the induced 2d metric. To see this, let us consider 
the normal 1-form $l_\m=-\p_\m u $. Since it is null, it is automatically geodesic and twist-free; and since the level sets of $u$ provide a null foliation, it is affinely parametrised. 
The associated congruence tensor coincides then with the Lie derivative of the induced metric, 
which in partial Bondi gauge is proportional to the radial derivative,

\be\label{NGC}
\na_A l_B = \f12 \pounds_l \g_{AB} = \f1{2N}\p_r \g_{AB}. 
\ee
This surface tensor can be familiarly decomposed into shear $\s_{AB}$ and expansion $\th$ as the trace-less and trace parts, 
\be
\f12 \pounds_l \g_{AB} =\f{\sqrt{\g}}2 \pounds_l \check\g_{AB} + \f12 \g_{AB} \pounds_l\ln\sqrt{\g} = \s_{AB} + \f12\g_{AB}\th.
\ee
Hence, the shear of the null congruence carries the same information of the conformal 2d metric, up to zero modes lost in the derivative and which are part of the corner data. 
The fact that (the bulk of the) constraint-free data can be described in terms of shear will allow us to easily identify them in the first-order formalism, where $\na_\m$ is an affine connection.

Here we used the Bondi gauge in order to identify the tangent vector field to the null geodesic congruence with a coordinate vector, thus simplifying Lie derivatives. A 2d space-like metric in $\Si$, its Lie derivative defining a shear, and associated Sachs' propagating equations, can be identified without this gauge-fixing: it suffices to use a $2+2$ decomposition, either in terms of two scalar fields defining a $2+2$ foliation (one being $u$), or in terms of a null dyad (one element being $l_\m$), as we will review below. The role of the gauge-fixing is nonetheless crucial to specify the explicit integration scheme of the constraints and the other field equations. Hence, it is possible to talk about physical degrees of freedom in a completely covariant way, as often done in the literature, although only once the gauge is completely fixed one can truly identify constraint-free initial data.

\subsection{Hamiltonian structure}\label{SecH}
The fact that the constraint-free data can be either described by the metric or the shear, its null-radial derivative, captures a well-known property of field theories on the light cone: the momentum conjugated to the fields does not depend on velocities, but on the null radial derivative of the field. Consider for instance a scalar field in Minkowski spacetime. Defining $x^\pm:=t\pm r$, and choosing $x^+$ as `time' for the canonical analysis, the conjugate momentum is
\be\label{LCc}
\pi(x^-,x^A) := \f{\d {\cal L}}{\d \dot\phi} = \p_-\phi(x^-,x^A),
\ee
where $A=2,3$ are the transverse coordinates. The independence of the momentum from the velocities gives rise to a primary constraint $\Phi:=\pi-\p_-\phi$, which is second class with itself, up to zero modes, see e.g. \cite{IoSergeyNull}. In the following, we will refer to this constraint as light-cone condition.
This fact, which is just a direct consequence of the fact that the normal vector to a null hypersurface is tangent to it, means that the momentum is not an independent variable, and can then be eliminated from the phase space. 
The physical phase space has thus $\infty^1$ dimensions per degree of freedom, instead of $\infty^2$ as in the space-like formulation, and the fields satisfy Dirac brackets defined by a suitable regularisation of $\p_-^{-1}$. Since we are not interested in this paper in the subtle infrared issues and boundary conditions, let us content ourselves to describe the symplectic structure of the theory looking at the symplectic potential. To that end, one can use the covariant phase space method (see e.g. \cite{AshtekarReula}), and read the symplectic potential from the variation of the action in presence of a null boundary. Consider for simplicity a free scalar field, and a null boundary given by a single light-cone $\Si$ ruled by $x^-$. Then the variation of the action gives the following boundary contribution,
\be\label{ThPhi}
\Th = \int_\Si \p_- \phi \, \d\phi.
\ee
This symplectic potential shows that the conjugate momentum to $\phi$ satisfies the light-cone condition  \Ref{LCc}, and announces the presence of $\p_-^{-1}$ in the Dirac bracket among the $\phi$'s.

The same structure arises in gauge theories (see e.g. \cite{Grange:1998gy}) and linearised general relativity around Minkowski \cite{Scherk:1974zm, TorreLin87}: the physical phase space has $\infty^1$ dimensions for each physical degree of freedom (a transverse mode in these examples), and the conjugate momentum is given by the null radial derivative of the mode itself.
Remarkably, it is also true in full, non-linear general relativity, with the momentum given by the shear, again a null radial derivative of the physical degrees of freedom as shown in \Ref{NGC}. 
The Hamiltonian analysis of general relativity on a null hypersurface has been performed in \cite{Torre:1985rw} using the $2+2$ formalism of \cite{dInverno:1980kaa}. Starting with a covariant kinematical phase space of canonical variables $(g_{\m\n}, \Pi^{\m\n}:=\d{\cal L}/\d\p_u g_{\m\n}$), one finds 6 first class and 6 second class constraints, for a resulting 2-dimensional physical phase space, as expected. The six first class constraints split in 3 hypersurface diffeomorphism generators plus three primary constraints imposing the vanishing of the conjugate momenta to the chosen shift vectors. 
The six second class are: the null hypersurface condition $g^{00}=0$, which in turns gauge-fixes the Hamiltonian constraint and makes it second class;\footnote{Up to zero modes: Measure-zero `parallel' time diffeomorphisms are still allowed. For instance, these contain the BMS super-translations \cite{BMS} for asymptotically flat spacetimes.} two light-cone conditions, the non-linear version of \Ref{LCc}; the vanishing of the momentum conjugated to the lapse $\cN$, and the vanishing of $\p_u g^{00}$.\footnote{
This last constraint may look puzzling. The problem is that imposing $g^{00}$ strongly in the action would lead to a variational principle missing one of the Einstein's equations. To avoid this `missing equation', the Hamiltonian in \cite{Torre:1985rw} is first constructed with arbitrary $g^{00}$, and $g^{00}=0$ is later imposed as initial-value constraint on the phase space. The additional constraint $\p_u g^{00}=0$ then simply arises as a secondary constraint preserving the first one under evolution. 
As explained in \cite{IoSergeyNull}, an advantage of working with a first order formalism is that one does not need this somewhat artificial construction: we can impose the gauge-fixing condition strongly in the action and still have a complete well-defined variational principle, thanks to the appearance in the action of the variable canonically conjugated to $g^{00}$. Furthermore, the on-shell value of the Lagrange multiplier for $g^{00}=0$, which is fixed by hand in  \cite{Torre:1985rw}, comes up dynamically as a multiplier equation.}

The  analysis of \cite{Torre:1985rw} is general and does not require the Bondi gauge:
we introduce a $2+2$ foliation by two closed 1-forms, $n^\a = d\phi^\a$ locally, with $\a=0,1$, normals to a pair of hypersurfaces. 
Instead of lapse and shift, we have two shift vectors and a `lapse matrix' $N_{\a\b}$, with inverse $N^{\a\b} := n^\a_\m n^{\b\m}$, 
and dual basis $n_\a^\m := g^{\m\n} N_{\a\b} n^\b_\n$. The only gauge-fixing is to take a null foliation defined say by the level sets of $\phi^0$, so that $N^{00}=0=N_{11}$, and the lapse (i.e. the Lagrange multiplier of the Hamiltonian constraint) turns out to be the off-diagonal component, $N_{01}=-\cN$.\footnote{The sign we use in this definition is opposite to the one of \cite{Torre:1985rw}, to match with our earlier choice $N>0$.} 
The induced space-like metric on the 2-dimensional surface orthogonal to both normals is then $\g_{\m\n}=g_{\m\n} - N_{\a\b} n^\a_\m n^\b_\n$.
In this formalism, we can identify covariantly the two physical degrees of freedom with $\check{\g}_{\m\n}$; their propagating equations as the two components of the Einstein equations obtained from the trace-less projection onto the 2d surface; and their Hamiltonian counterpart as the multiplier equations arising from the stabilisation of the two light-cone conditions. 

If we adapt the null coordinate, $\phi^0=u$, we have $n_1^\m=N_{01}n^{0\m}=\cN l^\m$. Unlike $l^\m$, $n_1^\m$ has non-vanishing affinity, given by $k_{(n_1)} = \pounds_{n_1}\ln\cN$, and its shear and expansion are $\cN$ times those of $l^\m$.
The partial Bondi gauge corresponds to putting to zero one of the two shift vectors, and only in this gauge the coordinate vector $\p_{\phi^1}$ is tangent to the null geodesics on $\Si$. 
As discussed above, the gauge-fixing is convenient for many reasons, principally to provide the explicit integration scheme of the Einstein's equations, in particular solving the constraints. Another advantage is that due to the presence of complicated second class constraints, it is difficult to  write the explicit Dirac bracket for the physical phase space. Gauge-fixing gets rid of gauge quantities and simplifies this problem. It becomes for instance straightforward to write the symplectic potential purely in terms of physical data.
For our purposes, we specialise here the analysis of \cite{Torre:1985rw} to the partial Bondi gauge, adapting coordinates so that $\phi^0=u$ and requiring $g^{0A}=0$, but keeping $r$ unfixed as to see explicitly the role of lapse and $\sqrt{\g}$. This partial gauge-fixing eliminates various gauge fields from the phase space, and one can isolate the induced 2d metric $\g_{AB}$ and its conjugate momentum density, which turns out to be 
\be\label{PiTorre}
\hat\Pi_{AB} := \sqrt{\g}\,\Pi^{AB} = \f{\d {\cal L}}{\d \dot\g_{AB}} 
= \f{\sqrt{\g}}2(\g_{AB}\g_{CD} - \g_{AC}\g_{BD})\pounds_{n_1}\g^{CD} - \sqrt{\g}\,\g_{AB}( \pounds_{n_1}\ln\cN +\f1{2\cN} \pounds_{n_0} N^{00}),
\ee
in terms of the dual basis $(n_0,n_1)$ defined above.
 Taking the trace-less and trace parts, it is immediate to identify them as the shear and expansion of the null-geodesic congruence of $n_1$,
\begin{align}\label{Pishear}
& \Pi_{AB}-\f12\g_{AB}\Pi = \f{\sqrt{\g}}2 \pounds_{n_1} \check\g_{AB}  = \s_{(n_1)AB}, \\
& \Pi :=  \g^{AB} \Pi_{AB} = - \th_{(n_1)} - 2k_{(n_1)} -\f1{\cN} \pounds_{n_0} N^{00}.
\end{align}
The first equation above is precisely the light-cone condition \Ref{LCc} for non-linear gravity: 
the two physical momenta are the null radial derivatives of the two physical modes of the metric, namely, the shear of $n_1$. 
The second equation shows that the trace of the momentum does not carry any additional information, although this may require a few words: first, the expansion can be determined from the dynamical fields (up to boundary values) using the Raychaudhuri equation; the lapse can always be fixed to $1$ with a radial diffeomorphism as mentioned above, thus removing the non-affinity term;\footnote{Canonically, the fact that changing $r$ can be used to fix $\cN=1$ follows from the fact that lapse transforms under radial diffeos like the radial component of a tangent vector. The alternative gauge-fixing, $r$ areal coordinate with lapse free, turns the non-affinity term into a corner contribution to the symplectic potential, see e.g. \cite{MikeNull}. As mentioned above, we do not discuss corner terms in the present paper.} finally, the last term vanishes using the equations of motion.

In this partial Bondi gauge, the symplectic potential computed in \cite{Torre:1985rw} reads\footnote{As usual, deriving the symplectic potential requires an integration by part. Although \cite{Torre:1985rw} does not give the associated boundary term, this is known to be $2 \int_\Si (\th+k)\sqrt{\g}$, see e.g. \cite{Parattu:2015gga}. Note the different factors of 2 between the boundary term and the symplectic potential.}
\be\label{Th1}
\Th = \int_\Si d^3x \ \hat\Pi^{AB}\d \g_{AB} 
= -\int_\Si d^3x \left[ \s_{(n_1)AB}\d\hat\g^{AB} + (\th_{(n_1)}+2k_{(n_1)})\d\sqrt{\g} \right],
\ee
where we used $\d\g_{AB} = -\g_{AC}\g_{BD}\d\g^{BD}$ 
and defined the densitised inverse metric $\hat{\g}^{AB}:=\sqrt{\g}\g^{AB}$.
Notice also that the shear term can be rewritten using $-\s_{AB}\d\hat \g^{AB} = \sqrt{\g}\, \s^{AB}\d \g_{AB}$.
The non-affinity term vanishes if we fix a constant lapse, and using the explicit metric form of shear and expansion, the symplectic potential takes the form 
\be\label{Th2}
\Th = -\int_\Si d^3x \left[  \f{\sqrt{\g}}2\pounds_{n_1} \check\g_{AB}  \d\hat\g^{AB} + \pounds_{n_1}\ln\sqrt{\g} \, \d\sqrt{\g} \right].
\ee
The first term has precisely the form \Ref{ThPhi} for the 2 physical degrees of freedom, which is the main point we wanted to make. The second term is just a corner contribution thanks to the Bondi gauge. In this paper we are interested in bulk degrees of freedom, hence we neglect corner terms in the symplectic potential.

This symplectic potential for the shear, here adapted from \cite{Torre:1985rw} to the Bondi gauge, can also be derived with covariant phase space methods (see e.g. \cite{AshtekarReula}), without referring to a special coordinate system but only to the field equations.
It plays a crucial role in the study of BMS charges at null infinity (see e.g. \cite{Ashtekar:2014zsa,Wald:1999wa,Barnich:2011mi}), which has recently received much attention for its possible relation to the information black hole paradox argued for in \cite{Hawking:2016sgy}.
For a careful treatment of caustics, corner data and residual diffeos, see \cite{Reisenberger:2007ku,MikeNull}, as well as \cite{Duch:2016kxr} in a related context. 
For a more general expression of $\Th$ without a full foliation and a discussion of corner terms without any coordinate gauge fixing, and its relevance to capture the full information about the charges, see \cite{FreidelNull}. See also \cite{Parattu:2015gga,Lehner:2016vdi,Jubb:2016qzt,Wieland:2017zkf}  for additional discussions on corner terms.

\section{Canonical structure in real connection variables}

\subsection{Tetrad and foliation}\label{Sec31}
In this section we briefly review the canonical structure of general relativity in connection variables on a null hypersurface \cite{IoSergeyNull}.
In units $16\pi G=1$, we work with the Einstein-Cartan action
\be
S[e,\om]=\f12\int_\mathcal{M}\eps_{IJKL} e^I\wedge e^J \wedge \left(F^{KL}(\omega)-\frac{\Lambda}{6}\, e^K\wedge e^L\right),
\label{action}
\ee
where $e^I$ is the tetrad 1-form, and
$F^{IJ}(\omega)=d\omega^{IJ}+{\omega^{I}}_K\wedge \omega^{KJ}$ the curvature
of the spin connection $\om^{IJ}$.
As in the ADM formalism, we fix a $3+1$ foliation with adapted coordinates $x^\m = (t,x^a)$, and hypersurfaces $\Si$ described by the level sets of $t$. 
We parametrise the tetrad as follows \cite{Barros, Alexandrov:1998cu,AlexandrovSO4cov},
\be
e^0=\hat N d t+\chi_i E_a^i d x^a,
\qquad
e^i=N^a E_a^i d t +E_a^i d x^a.
\label{e}
\ee
The hypersurface normal is then the soldering of the internal direction $x_+^I:=(1,\chi^i)$: 
\be\label{nS}
n^\Si_\m := e_\m^I x_{+I}=(-N,0,0,0), \qquad \cN = \hat N - N^a E^i_a\chi_i.
\ee
For space-like $\Si$, the usual tetrad adapted to the ADM coordinates is recovered for vanishing $\chi^i$, which makes $e^0$ parallel to the hypersurface normal.
Using a non-adapted tetrad may appear as an unnecessary complication, but has the advantage that allows one to control the nature of the foliation.
The metric induced by \Ref{e} on $\Si$ is
\be
q_{ab} :=e_a^Ie_b^J\eta_{IJ}= {\cal X}_{ij} E_a^iE_b^j, \qquad  {\cal X}_{ij} := \delta_{ij}-\chi_i\chi_j, \qquad \det q_{ab} = E^2 (1-\chi^2),
\label{projchi}
\ee
where $\chi^2:=\chi_i\chi^i$. It is respectively space-like for $\chi^2<1$, null for $\chi^2=1$, and time-like for $\chi^2>1$.
In other words, we control with $\chi^i$ the signature of the hypersurface normal, while $e^0$ is always time-like.

We are interested here in the case of a foliation by null hypersurfaces.
Notice that even though the induced hypersurface metric is degenerate,  
we can still assume an invertible triad, with inverse denoted by $E^a_i$. 
This means that we can use the triad determinant, $E:=\det E_a^i\neq 0$, to define tensor densities.
We denote such densities with a tilde respectively above or below the tensor, e.g. $\tE^a_i:=EE^a_i$ for density weight 1 and $\Et_a^i:=E^{-1}E_a^i$ for density weight $-1$.
The triad invertibility is an advantage of the tetrad formalism for null foliations, and it 
allows us to write the null direction of the induced metric on $\Si$ as $(E^a_i\chi^i)\p_a$. Further, although the induced metric $q_{ab}$ is not invertible, we can raise and lower its indices with the triad. We define the projector $q^a{}_b:=E^{ai}E_b^j{\cal X}_{ij}$, which projects hypersurface vectors on 2d space-like spaces orthogonal to the null direction $E^a_i\chi^i$; and $q^{ab}:= E^a_iE^b_j{\cal X}^{ij}$, which satisfies $q^{ab}q_{bc} = q^a{}_c$. 

On the other hand, $\hat N$ and $N^a$ should not be immediately identified with the lapse and shift functions, defined as the Lagrange multipliers of the diffeomorphism constraints.
The true lapse can be identified from \Ref{nS} or by computing the tetrad determinant, which turns out to be 
$e=\cN E$. As for the shift vector, there is no canonical choice on a null foliation, corresponding to the fact that there is no canonical Hamiltonian.\footnote{In the sense that it is not possible to express the Hamiltonian constraint purely in terms of hypersurface data, see for instance \cite{IoSergeyNull} and \cite{Torre:1985rw}.} Following the canonical analysis of \cite{IoSergeyNull}, to be recalled below, we keep $N^a$ as the shift vector. 
In terms of the lapse $\cN$, the metric associated with the tetrad \Ref{e} reads
\be\label{g}
g_{\mu\nu}=\left(
\begin{array}{cc}
-\cN^2 +N^a N^b q_{ab}-2\cN N^a E_a^i\chi_i & q_{bc}N^c -\cN E_b^i\chi_i
\\
q_{ac}N^c-\cN E_a^i\chi_i & q_{ab}
\end{array}
\right),
\ee
with inverse
\be\label{ginv}
g^{\mu\nu}= \f1{\cN} \left(\begin{array}{cc}
0 & - E^b_i\chi^i
\\
- E^a_i\chi^i & \cN E^a_i E^b_i+ (N^aE^b_i+N^bE^a_i)\chi_i
\end{array}\right).
\ee
The coordinate $t$ being adapted to the null foliation, $g_{ab}\equiv q_{ab}$ is the degenerate induced metric on $\Si$. We can also write the projector on the 2d space-like spaces in a covariant form, using the null dyad provided by the
internal null vectors $x_{\pm}^I=(\pm 1,\chi^i)$ soldered by the tetrad,
\be\label{xpm}
x^I_{\pm} := (\pm 1, \chi^i), \qquad x_{\pm\m} = e^I_\m x_{\pm I} = \left\{ \begin{array}{l}
(-\cN, 0) \\ (\cN+2N^aE_a\chi, 2E_a\chi) \end{array} \right., \qquad x_{+\m} x_-^\m=2. 
\ee
We then have
\be\label{perp}
\perp^\m{}_\n:=\d^\m{}_\n -\f12 x_+^\m x_{-\n}-\f12 x_-^\m x_{+\n} =  \mat{0}{0}{q^a{}_{b}N^b}{q^a{}_{b}},
\ee
and 
\be\label{gind}
\g_{\m\n} := g_{\m\n} - x_{+(\m} x_{-\n)} = \mat{q_{ab} N^a N^b}{q_{bc}N^c}{q_{ab}N^b}{q_{ab}}
\ee
is the induced metric in covariant form. For later purposes, let us identify here the propagating Einstein's equations, which are given by the components
\be\label{dynEqs}
(\perp {\bG}^{\rm T})^{ab} := \Big(\perp^a{}_{(\r}\perp^b{}_{\s)}-\tfrac12\perp^{ab}\perp_{\r\s}\Big) \bG^{\r\s}
= \Pi^{ab}_{cd} \Bigl( \bG^{cd}+N^d \bG^{c0}+N^{c} \bG^{0d}+N^c N^d \bG^{00}\Bigr).
\ee
Here
\be
\Pi^{ab}_{cd} := q^{a}_{(c} q^{b}_{d)}- \f12\, q^{ab} q_{cd}
\label{projPi}
\ee
is the traceless part of the projector on $S$ for symmetric hypersurface tensors,
and we used the notation $\bG^\m_I:= G^\mu_I + \L e^\m_I = 0$ where $G^\m_I$ is the Einstein tensor in tetrad indices.
The explicit form of \Ref{dynEqs} is given in \cite{IoSergeyNull}, and it will not be needed here. 

An advantage of the tetrad formulation is that we can perform the canonical analysis with the $3+1$ null foliation \cite{IoSergeyNull}, without the need of  introducing a further $2+2$ foliation like in the metric case. 
Nonetheless, it is instructive to review how the two formalisms compare in the absence of torsion. 
Our coordinates are adapted to the $3+1$ foliation by null hypersurfaces with normal 1-form $dt$, and to match notations with the literature, we rename from now on $t=u$; however the 2d space-like spaces defined by \Ref{perp} are in general not integrable, hence they do not foliate spacetime. Nonetheless, we can choose a $2+2$ foliation and adapt our tetrad to it. For the sake of simplicity let us choose the foliation given by the normals 
\be
n^0=du, \qquad n^1=dr,
\ee
so that our coordinates $x^a=(r,x^A)$ are already adapted, 
and the induced 2d metric is $\g_{AB}\equiv g_{AB}=q_{AB}$.
To adapt the null dyad $x_{\pm\m}$ to this foliation we use the translational part of the ISO(2) group stabilising $x_+^I$ to remove the components $x_{-A}=E_A^i\chi_i=0$. This gauge transformation makes the tangent vectors to $\{S\}$ integrable. The same can be done in the Newman-Penrose formalism, see Appendix~\ref{Sec2+2} for details and a general discussion.
Comparing then the metric coefficients of \Ref{gS} and \Ref{ginv} we see that the lapse functions used in the metric and connection formulations differ by a factor $E^r_i\chi^i$. This can be always set to one with an internal boost along $x_+^I$, as explained in the next Section. Hence, using this boost and the translational part of the stabiliser we can always reach the internal `radial gauge' 
\be\label{nullgauge}
E_{a}^i\chi_i=(1,0,0) \quad \Leftrightarrow \quad E^r_i\chi^i=1, \ {\cal X}^{ij}E^r_j=0,
\ee
where the equivalence follows from the invertibility of the triad. 
In this internal gauge $N$ coincides with the lapse of the metric formalism, given by $-1/g^{01}$ in adapted coordinates, $E=\sqrt{\g}$ and $\sqrt{-g}=NE=N\sqrt{\g}$, and the induced metrics coincide, $g_{\m\n} - x_{+(\m} x_{-\n)} = g_{\m\n} - N_{\a\b} n^\a_{\m} n^\b_{\n}$. 
Proofs and more details on the relation between the $\chi$-tetrad and the 2+2 formalism are reported in Appendix~\ref{Appchi}.

\subsection{Constraint structure}\label{Sec32}
On a null hypersurface, each degree of freedom is characterised by a single dimension in phase space, as recalled above. This means that the constraint structure associated to the gravitational action should lead to a phase space of dimensions $2\times\infty^3$ on $\Si$ (plus eventual zero modes and corner data, not discussed here). We now review from \cite{IoSergeyNull} how this counting comes about, as the result has some peculiar aspects that we wish to analyse in this paper.

From \Ref{action}, we see that the canonical momentum conjugated to $\om^{IJ}_a$ is $\tP^a_{IJ}:=(1/2)\eps^{abc}\eps_{IJKL} e_b^K e_c^L$, namely, it is simple as a bi-vector in the internal indices. This results in a set of (primary) simplicity constraints, which fixing an internal null direction,
can be written in linear form as $\Phi^a_I:={\eps_{IJ}}^{KL}\tP^a_{KL} x_+^J=0$. 
Two different canonical analysis were presented in \cite{IoSergeyNull}. The first is manifestly covariant, with only $\chi^2=1$ as a gauge-fixing condition. The second gauge-fixes instead all three components, that is $\chi^i=\hat\chi^i$ for a fixed $\hat\chi^i$ with $\hat\chi^2=1$. Since in this paper we are interested in the identification of constraint-free data that arises through a complete gauge-fixing, we recall only the details of the second analysis, and refer the reader interested in the covariant expressions to \cite{IoSergeyNull}.

Working with a gauge-fixed internal direction, we can solve explicitly the primary simplicity constraints in terms of $\tP^a_{0i}=\tE^a_i, \tP^a_{ij}=2\tE^a_{[i}\chi_{j]}$. The kinetic term of the action is then diagonalised by the same change of connection variables as in the space-like case \cite{Alexandrov:1998cu},
\be
\omega_a^{0i}= \eta_a^i-\omega_a^{ij}\chi_j,
\qquad
\omega_a^{ij}=\,\eps^{ijk} \left(\tl r_{kl}+\f12 \eps_{klm}\tl\om^m\right)\Et_a^l,
\label{decomp-con}
\ee
with $\tl r_{ij}$ symmetric.
After this change of variables and an integration by parts, the action reads\footnote{In \cite{IoSergeyNull} we rescaled the action by a factor $1/2$, to avoid a number of factors of 2 when computing Poisson brackets. Here we restore the conventional units. Accordingly, the parametrization of $\tP^a_{IJ}$ in terms of $\tE^a_i$, as well as the explicit expressions for the constraints presented below in \Ref{primary}, are twice those of \cite{IoSergeyNull}.}
\be
S=
\int dt \int_\Si 2(\tE^a_i\p_t \eta_a^{i}+\pi^{ij} \p_t \tl r_{ij} + \chi_i\p_t \tl \om^i)
+\l_{ij} \Phi^{ij} + \m_i \vphi^i
+n^{IJ} {\cal G}_{IJ} + N^a{\cal D}_a +\tNn \hh,
\label{actionH}
\ee
where
\be
{\cal G}_{IJ}:= \, D_a\tP^a_{IJ},
\qquad {\cal D}_a:=\,-\tP^b_{IJ} F_{ab}^{IJ}+\om^{IJ}_a {\cal G}_{IJ},
\qquad \hh:=\, \tE^a_i\tE^b_j F_{ab}^{ij} - 2 {\Lambda} E^2,
\label{primary}
\ee
are the gauge and diffeomorphism constraints, written in covariant form for practical reasons. Notice that as in the space-like case, the generator of spatial diffeomorphism includes internal gauge transformations (and accordingly, we have $n^{IJ} = \om_0^{IJ} - N^a \om_a^{IJ}$). Next, the constraint
\be
\Phi^{ij} = \pi^{ij}
\ee
imposes the vanishing of the momentum conjugated to $r^{ij}$, and is the left-over of the primary simplicity constraints in this non-covariant analysis.
Finally, the constraint
\be\label{Cphi}
\vphi^i = \chi^i-\hat\chi^i
\ee
gauge-fixes the internal vector. In particular, the projection $(\chi_i+\hat\chi^i)\vphi^i$ gives the null-foliation condition $\chi^2=1$, namely $g^{00}=0$, and its stabilisation plays an important role in recovering all of Einstein's equations.\footnote{This plays the role of the $\p_ug^{00}=0$ condition of \cite{Torre:1985rw}, and the advantage of the first-order formalism is that it can be imposed prior to computing the Hamiltonian.}

The phase space of the theory has initially 36 dimensions, with Poisson brackets
\begin{align}\label{PBe}
& \{\eta_a^i(x), \tE^b_j(x')\} = \f12 \d^i_j \d^b_a \d^{(3)}(x,x'), \\
\nonumber & \{\tl r^{ij}(x), \pi_{kl}(x')\} = \f12 \d^{ij}_{(kl)} \d^{(3)}(x,x'), \qquad \{\tl\om^i(x), \chi_j(x')\} = \f12 \d^i_j \d^{(3)}(x,x').
\end{align}
The explicit form of the constraints is considerably more compact and elegant than in the metric case \cite{Torre:1985rw}, a fact familiar from the use of Ashtekar variables in other foliations. On the other hand, many of the constraints are second class.
The reader familiar with the Hamiltonian analysis in the space-like case will recall that the stabilisation of the primary simplicity constraints leads to six secondary constraints which are second class with the primary. The secondary constraints thus obtained, together with the six Gauss constraints, recover half of the torsion-less conditions; the remaining half goes in Hamiltonian equations of motion.
In the null case the situation becomes more subtle: there are again six secondary constraints, given by
\be
\Psi^{ij} 
=-\eps^{(ikl}\tE^a_k\Et_b^{j)} \p_a\tE^b_l+\eps^{(ikl}\tE^a_k\chi_l\eta_a^{j)}
-{\cal M}^{ij,kl} r_{kl},
\label{noncovPsi}
\ee
where
\be
{\cal M}^{ij,kl}=\eps^{(ikm}\eps^{j)ln}{\cal X}_{mn}.
\label{matrMij}
\ee
These have the same geometric interpretation of being six of the torsion-less conditions. However, only four of them are now automatically preserved. This is a consequence of the fact that \Ref{matrMij} has a two-dimensional kernel: $\Pi^{ij}_{kl} {\cal M}^{kl,mn}\equiv 0$,
where $\Pi^{ij}_{kl}$ is the internal version of the symmetric-traceless projector \Ref{projPi} obtained via the triad.
Then, stabilisation of the two secondary constraints
\be\label{Psi2}
\hat \Psi^{ij} = \Pi^{ij}_{kl} \Psi^{kl},
\ee
requires two additional, tertiary constraints
\be
\Upsilon^{ab} := \f12 \,\Pi^{ab}_{cd} E^{(c}_i\eps^{d)ef}\left( F^{0i}_{ef}- \chi_j F^{ij}_{ef}\right)=0.
\label{tertcon}
\ee

As pointed out in \cite{IoSergeyNull}, the two constraints \Ref{Psi2} are the light-cone conditions imposing the proportionality of physical momenta to the hypersurface derivatives in the null direction: As we will show below, they reproduce precisely the metric relation  \Ref{Pishear} between momenta and shear. 
What is peculiar to the formalism, is that this condition is not automatically preserved under the evolution, but requires the additional constraints \Ref{tertcon}. These additional constraints are not torsion-less conditions; they will be discussed in details in Section \ref{SecTert} below. 

Concerning the nature of the constraints and the dimension of the reduced phase space, we have the following situation. The hypersurface diffeos ${\cal D}_a$ are first class, but not the Hamiltonian $\cal H$, which forms a second class pair with $\chi_i\vphi^i$. The other two components ${\cal X}_{ij}\vphi^j$ gauge-fix two of the six Gauss constraints, those that would change the internal direction. The other four Gauss constraints remain first class. This is different from the canonical analysis on a space-like or time-like hypersurface, where fixing the internal direction gives a 3-dimensional isometry group. Here instead we have a 4-dimensional isometry group, given by the little group ISO(2) of the internal direction given by $\chi^i$, plus boosts along $\chi^i$. The fact that the isometry group on a null hypersurface is one dimension larger than for other foliations is of course a well-known property, that led Dirac himself to suggest the use of null foliations as preferred ones. In the context of first-order general relativity with complex self-dual variables, it has for instance been pointed out in \cite{Goldberg:1992st,dInverno:2006mmy}. 

However, there is a subtle way in which this extra isometry is realised in our context, because the action of internal boosts along $\chi^i$ mixes with that of radial diffeomorphisms. Let us spend a few words explaining it.
Notice that right from the start we fixed to unity the $0$-th component of the internal null direction $x_+^I$. This choice, implicit in the parametrization \Ref{e} of the tetrad, deprives us of the possibility of changing $\chi^i$ with an internal boost along $\chi^i$, since in the absence of a variable $x_+^0$ this would not preserve the light-likeness of the internal direction. Nonetheless, the explicit calculation of the constraint structure shows that $K_\chi:={\cal G}_{0i}\chi^i$ is still a first class constraint: simply, its action is not to change $\chi^i$, which it leaves invariant, but rather to rescale the lapse function. Using the transformation properties for Lagrange multipliers (see e.g. \cite{HenneauxBook}), we find for the smeared constraint the transformation
\be\label{deltaN}
K_\chi(\l) \triangleright \cN = e^\l \cN.
\ee
In other words, the lapse function is in our formalism soldered to the extent 
of the internal null direction, see \Ref{xpm}, and this is the reason why it transforms under internal radial boosts. As already discussed at the end of Sec.~\ref{Sec31}, our lapse coincides with the lapse of the metric formalism only if we fix the radial boosts to have $E^r_i\chi^i=1$.
Hence, there is in our formalism a partial mixing of the action of internal boosts along $\chi^i$ and radial diffeomorphisms.

To complete the review of the constraints structure, it remains to discuss the simplicity constraints. They are all second class, but in different ways: $\hat\Psi^{ij}$ among themselves, just like those encoding the light-cone conditions \Ref{LCc}, the remaining four $\Psi^{ij}$ are second class with four of the primary $\Phi^{ij}$, and the remaining two $\Phi^{ij}$ are second class with the two tertiary constraints.
The overall canonical structure established in \cite{IoSergeyNull} leads to the following diagram, where the arrows indicate which constraints are mutually second class:
\vspace{0.4cm}
\begin{center}
\begin{tabular}{cccccc}
primary constraints & $\Phi^{ij}$ & $\vphi^i$ $\stackrel{2}{\leftrightarrow}$ ${\cal G}_{IJ}$ & ${\cal D}_a$ & $\hh$&
\\
& $\updownarrow\scriptstyle{4}$ & & & &
\\
secondary constraints & $\Psi^{ab}$ & && &
\\
& $\displaystyle{\circlearrowright\scriptstyle{2}}\atop \vphantom{a}$ & & & &
\\
tertiary constraints & $\Upsilon^{ab}$& & & &
\end{tabular}
\end{center}

\vspace{-3.5cm}\hspace{5.2cm}
\unitlength 0.44mm 
\linethickness{0.4pt}
\ifx\plotpoint\undefined\newsavebox{\plotpoint}\fi 
\begin{picture}(110,74)(0,0)
\put(57,7){\vector(-3,-4){.09}}\put(57,49){\vector(-3,4){.09}}\qbezier(57,49)(73,28)(57,7)
\put(71,52){\vector(-4,-3){.09}}\put(132,52){\vector(4,-3){.09}}
\qbezier(72,53)(101,71)(131,53)
\put(69,30){\makebox(0,0)[cc]{\scriptsize 2}}
\put(101,58){\makebox(0,0)[cc]{\scriptsize 1}}
\end{picture}\label{page-diag}
\vspace{0.3cm}
\\
We have 7 first class constraints (forming a proper Lie algebra), and 20 second class constraints, for a \linebreak $2\times \infty^3$-dimensional physical phase space, as expected for the use of a null hypersurface. Among those, the pair Hamiltonian-null hypersurface condition.

\section{Geometric interpretation}
\subsection{Newman-Penrose tetrad}

To elucidate the geometric content of the canonical structure in the first order formalism, it is convenient to use the Newman-Penrose (NP) formalism. To that end, we want to map our tetrad \Ref{e} to a doubly-null tetrad  $(l^\m,n^\m,m^\m,\bar m^\m)$, where
\be
l_\m n^\m = -1 = -m_\m \bar m^\m, \qquad g_{\m\n} = -2l_{(\m} n_{\n)} + 2 m_{(\m} \bar m_{\n)}.
\ee
We have already partially done so, when we introduced the soldered internal null vectors $x_{\pm}^\m= e^\m_I x_{\pm}^I, \ x_{\pm}^I=(\pm 1,\chi^i)$, which provide the first pair. For the second pair, we have to choose a spatial dyad for the induced metric \Ref{gind}, that is $\g_{\m\n}=2m_{(\m}\bar m_{\n)}$; we can do so taking $m^\m$ to be a complex linear combination of the two orthogonal tetrad directions ${\cal X}^{ij}e_j^\m$, normalised by  $m_\m\bar m^\m=1$. The set 
\be\label{xNP}
(x_{+}^\m,-x_-^\m,m^\m,\bar m^\m)
\ee
so defined is an NP tetrad. Notice that $x_{+\m}=-N\p_\m u$, so the first vector chosen is normal to the null hypersurface. The minus sign in front of the second vector is to follow the conventions to have all vectors future-pointing.

Before adopting the traditional notation with $l^\m$ and $n^\m$ for the first two vectors, let us briefly discuss the frame freedom.
Using the nomenclature of \cite{Chandra}, we have rotations of class $I$ leaving $l^\m$ unchanged, of class $II$ leaving $n^\m$ unchanged, and of class $III$ rescaling $l^\m$ and $n^\m$ and rotating $m^\m$: 
\be\label{classIII}
l^\m \mapsto A^{-1} l^\m, \qquad n^\m \mapsto A n^\m, \qquad m^\m \mapsto e^{i\th} m^\m.
\ee
Conforming with standard literature on null hypersurfaces, we want the first null co-vector to be normal to the null hypersurface and future pointing, that is  $l_\m\propto -\p_\m u$. 
Concerning its `normalisation', a reasonable choice is to take it proportional to the lapse function, like in the space-like Arnowitt-Deser-Misner (ADM) canonical analysis: $l_\m^{\rm{\scriptsize ADM}}=-N\p_\m u$. This analogy with ADM is confirmed by Torre's analysis, which as we recalled above, identifies in $n_{1\m}
\equiv l_\m^{\rm{\scriptsize ADM}}$ the normal relevant to the Hamiltonian structure, namely whose shear gives the conjugate momentum in the action.
However, most of the literature on null hypersurfaces uses a gradient normal, $l_\m = -\p_\m u$, and we'll conform to that, by taking 
\be\label{Nres}
l^\m = \f1\cN x_+^\m, \qquad n^\m=-\f\cN2 x_-^\m. 
\ee 
This rescaling of $x_\pm^\m$ means paying off a large number of $N$ factors in the spin coefficients, see the explicit expressions reported in Appendix~\ref{AppSC}. In any case, the relation between the two choices is a class $III$ transformation, and all NP quantities are related by simple and already tabulated transformations that can be found in \cite{Chandra}, some of which are reported in Appendix~\ref{AppSC}.\footnote{The rescaling also means that while all Lorentz transformations of \Ref{xNP} are generated canonically via ${\cal G}_{IJ}$, this is not the case for $(l,n)$ defined via \Ref{Nres}: we disconnect the canonical action of the radial boost $K_\chi$, which leaves them invariant instead of generating the class $III$ rescaling. We see then again that $l^{\rm ADM}_\m=x_{+\m}$ is a more canonical choice of null tetrad adapted to the foliation.}

We fix from now on the following internal direction,
\be\label{chifixed}
\chi^i=(1,0,0),
\ee
and introduce the notation $v^\pm\equiv v_\pm :=\f1{\sqrt{2}} (v^2\pm iv^3)$ for the internal indices $M=2,3$ orthogonal to it. 
This choice is done only for the convenience of writing explicitly the tetrad components of $m^\m$ and $\bar m^\m$ when needed, and we will keep referring to $\chi^i$ in the formulas as to make them immediately adaptable to other equivalent choices.
Summarising, our NP tetrad and co-tetrad, and their expressions in terms of the metric coefficients \Ref{e}, are
\begin{subequations}\label{NP}\begin{align}
& l^\m=\f1\cN(e_0^\m+e_1^\m) = (0,\f1\cN E^a_i\chi^i), \\ 
& n^\m=\f{\cN}2(e_0^\m-e_1^\m) = (1,-N^a-\tfrac12{\cN} E^a_i\chi^i), \\ 
& m^\m = \f1{\sqrt{2}}(e_2^\m - ie_3^\m) = (0, E^a_-),
\end{align}\end{subequations}
and
\begin{subequations}\label{coNP}\begin{align}
& l_\m=\f1\cN(-e^0_\m+e^1_\m) = (-1,0), \\ 
& n_\m=-\f{\cN}2(e^0_\m+e^1_\m) = -\Big(\f N2(N+2N^aE_a^i\chi_i), NE_a^i\chi_i\Big), \\ 
& m_\m = \f1{\sqrt{2}}(e^2_\m - ie^3_\m) = (N^aE_a^-, E_a^-).
\end{align}\end{subequations}
The NP tetrad thus constructed is adapted to a null foliation like the one used in most literature \cite{Newman:1962cia,NewmanTod,Adamo:2009vu}. A detailed comparison and discussion of the special cases corresponding to a tetrad further adapted to a $2+2$ foliation or to the Bondi gauge can be found in Appendix~\ref{Sec2+2} and \ref{Appchi}.

Associated with the NP tetrad are the spin coefficients, namely 12 complex scalars projections of the connection $\om_\m^{IJ}$, e.g. (minus) the complex shear $\s:=-m^\m m^\n \na_\n l_\m$.\footnote{The reader familiar with the NP formalism will notice an opposite sign in this definition. This is a consequence of the fact that we work with mostly plus signature.} If the connection is Levi-Civita, these are functions of the metric. In the first order formalism on the other hand, the connection is an independent variable, and the NP spin coefficients will be functions of the metric and of the connection components. To distinguish the two situations, we will keep the original NP notation, e.g. $\s$, for the Levi-Civita coefficients, and add an apex $\circ$ for the spin coefficients with an affine off-shell connection, e.g. $\os{\s}$. On-shell of the torsion-less condition, $\om^{IJ}=\om^{IJ}(e)$ and $\os{\s}=\s$.
Explicit expressions for all the spin coefficients are in Appendix \ref{AppSC}, and we will report in the main text only those relevant for the discussion.

\subsection{The affine null congruence}\label{SecNC}

Since the normal vector $l^\m$ is null, it would be automatically geodesic with respect to the spacetime Levi-Civita connection. Furthermore it would have vanishing non-affinity since it is the unit normal to a null foliation. With an off-shell, affine connection $\om^{IJ}_a$ on the other hand, these familiar properties do not hold.
Using Newman-Penrose notation with an apex $\circ$ for the spin coefficients of the affine off-shell connection, what we have is
\be
l^\n\na_\n l^\m = 
\os{\eps}\, l^\m - \os{\k} \, \bar m^\m + {\rm cc},
\ee
with `non-affinity' and `non-geodesicity' that are given respectively by
\be\label{kk}
k_{(l)} := \os{\eps}+{\rm cc} = -\f1N E^a_i\chi^i (  \eta_a^i\chi_i - \p_a\ln\cN ), \qquad \os{\k} = - \f1{N^2} E^a_i\chi^i \, \eta_a^-.
\ee
For the same reason, the congruence $\na_\m l_\n$ is not twist-free, even though $l_\m$ is the gradient of a scalar, nor defined intrinsically on $S$:
it also carries components away from it.
Nonetheless, we can still take its projection $\perp^\r{}_\m \perp^\s{}_\n \na_\r l_\s$, and decompose it into irreducible components: we will refer to the traceless-symmetric $\os{\s}_{\m\n}$, trace $\os{\th}$ and antisymmetric parts $\os{\om}_{\m\n}$ as `connection shear', `connection expansion', and `connection twist'. 
The components away from the hypersurface $\Si$, which are all proportional to the shift vector $N^a$, are not directly relevant for us and we leave them to Appendix~\ref{AppCongruence}.
Using the definition $\na_\m e^I_\n = -\om^{IJ}_\m e_{\n J}$ and the decomposition \Ref{decomp-con}, we have for the hypersurface components
\be
\na_a l_b = \f1{\cN}{\cal X}_{ij} \eta^i_a E^j_b,
\ee
and
\be\label{lNGC}
\os{\s}_{(l)\,ab} := \f1{N} q_a{}^c q_b{}^d {\cal X}_{ij} \eta_{(c}^i E^j_{d)} - \f12 q_{ab} \os{\th}_{(l)}, \qquad \os{\th}_{(l)}:= \f1N {\cal X}_{ij} \eta_a^i E^a_j, 
\qquad \os{\om}_{(l)\,ab}:= \f1N q_a{}^c q_b{}^d {\cal X}_{ij} \eta_{[c}^i E^j_{d]}.
\ee
In NP notation, shear, twist and expansion are described by the following two complex scalars, 
\begin{align}
& \os{\s} := - m^\m m^\n \na_\n l_\m = - m^\m m^\n \os{\s}_{(l)\,\m\n} = - \f1N E^a_-\eta_a^-, \\
& \os{\r}:= - m^\m \bar m^\n \na_\n l_\m = -\f12 \os{\th}_{(l)} - m^\m\bar m^\n \os{\om}_{(l)\,\m\n} = -\f1N E^a_+\eta_a^-,
\end{align}
where the real and imaginary parts of $\os{\rho}$ carry respectively the connection expansion and twist.
It is also convenient to introduce the complex shear $\os{\s}_{(l)} := m^\m m^\n \os{\s}_{(l)\,\m\n} = - \os{\s}$. This comes up awkwardly opposite in sign to the NP spin coefficient, but the minus sign is an unavoidable consequence of the fact that we work with mostly plus signature, the opposite to NP.

The connection shear so computed allows us to identify Sachs' constraint-free initial data for first-order general relativity in terms of real connection variables: in the absence of torsion, $\os{\s}=\s$ and we can follow the same hierarchical integration scheme. From the connection perspective, the relevant piece of information is thus $E^a_-\eta_a^-$; namely the contraction with the triad of $\eta_a^-$, which is the translation part of the ISO(2) stabilising the null direction $x_+^I$.
Notice that both connection term and triad term have the same internal helicity: loosely speaking, it is this coherence that allows to reproduce the spin-2 behaviour in metric language.

Notice that at the level of Poisson brackets, the shear components commute: trivially in $\{\os{\s}_{(l)},\os{\s}_{(l)}\}=0$, but also when the conjugate appears, since\footnote{Using the brackets \Ref{PBe}, and notice that $\{\eta_a^i,\tE^b_j/(NE)\}=1/(2NE)(\d^b_a\d^j_i-E^a_i E^j_b/2)$.}
\be
\big\{\os{\s}_{(l)}, \bar{\os{\s}}_{(l)}\big\} = \f{2i}{NE} \im(\os{\r}),
\ee
which vanishes on-shell of the Gauss law, as we show in the next Section. This is to be expected, since it is only at the level of the Dirac bracket that the shear components do not commute with themselves, that is when the light-cone constraints are used. We will show below in Section \ref{SecSympl} that the Dirac bracket reproduces exactly the metric structure of \Ref{Th2}.

In terms of the covariant connection, the shear, twist and expansion are described as follows,
\be\label{scov}
\os{\s}_{(l)}= e_I^\n e_J^\r \,  m^\m m_\n l_\r \, \om_\m^{IJ}, \qquad
\os{\r}= e_I^\n e_J^\r \, \bar m^\m m_\n l_\r \, \om_\m^{JI}. 
\ee
Using these covariant expressions, it is easy to see how the congruence is affected by the presence of torsion, writing $\om^{IJ}_\m = \om^{IJ}_\m(e)+C^{IJ}_\m$ where $C^{IJ}_\m$ is the contorsion tensor. 
For instance, consider the case of fermions with a non-minimal coupling \cite{Alexandrov:2008iy}
\be
S_{\psi} = -\f i4\int e \bar \psi e^\m_I \g^I (a-i b\g^5) D_\m(\om) \psi + {\rm cc}, \qquad a,b\in\C, \qquad \re(a)\equiv 1.
\ee
(The minimal coupling would be $a=1$, $b=0$).
Solving Cartan's equation, one gets (restoring for a moment Newton's constant $G$)
\be
C^{IJ}_\m = 2\pi e_\m^K G \left[ \f12\eps^{IJ}{}_{KL}\Big( A^L - \im(b)V^L\Big) - \d_K^{[I} \Big(\re(b)A^{J]}+\im(a)V^{J]}\Big)\right], 
\ee
where $V^I=\bar\psi\g^I\psi$  and $A^I=\bar\psi\g^I\g^5\psi$ are the vectorial and axial currents. Plugging this decomposition into \Ref{scov} we find
\be
\os{\s}=\s, \qquad \os{\r}=\r-\pi G \Big[ i n_\m \Big( A^\m - \im(b)V^\m\Big) - l_\m \Big(\re(b) A^\m + \im(a)V^\m\Big) \Big].
\ee
The connection shear recovers its usual metric expression, whereas twist is introduced proportional to the axial current; for non-minimal coupling, the twist depends also on the vectorial current, and furthermore the expansion is modified, picking up an extra term proportional to the time-like component of the vectorial and axial currents.
More in general, for an arbitrary contorsion decomposed into its three irreducible components ${\bf (3/2,1/2)\oplus(1/2,3/2)\oplus(1/2,1/2)\oplus(1/2,1/2)}$,
\be
C^{\m,\n\r} = \bar{C}^{\m,\n\r} + \f23 g^{\m[\rho} \check{C}^{\n]} + \f1e \eps^{\m\n\r\s} \hat C_{\s},
\ee
we have
\be
\os{\s}=\s - m_\m m_\n l_\r \bar C^{\m,\n\r}, \qquad \os{\r}=\r-\bar m_\m m_\n l_\r \bar C^{\m,\n\r} +\f13 l_\m \check C^\m -i n_\m \hat C^\m,
\ee
as well as
\be
\os{\k} =\k - l_\m m_\n  l_\r \bar C^{\m,\n\r}, \qquad \os{k}_{(l)} = {k}_{(l)} - l_\m n_\n  l_\r \bar C^{\m,\n\r} - \f13 l_\m \check C^\m
\ee
for the non-geodesicity and inaffinity.

It is now instructive to see how the various quantities introduced above, and associated with an affine geodesic, are put on-shell by the constraints present in the Hamiltonian formulation of the theory, and thus (in the absence of torsion) take their values as in the more familiar metric formalism. As we show in details in the next subsection, the congruence is made geodesic by the Gauss law, which also puts on-shell the connection twist and expansion; the non-affinity vanishes as a consequence of the equation of motion stabilising $\chi^2=1$, namely the condition of null foliation; and finally, the connection shear is put on-shell by the two secondary simplicity constraints \Ref{Psi2}.\footnote{Notice that here we are defining the congruence in the presence of torsion using a displacement vector $\eta^\m$ such that $B_{\m\n}\eta^\n:=l^\n\na_\n \eta^\m=\eta^\n\na_\nu l_\m$. This is suggested as to keep the geometric interpretation of the spin coefficients $\os{\s}$ and $\os{\r}$, however it means that the displacement vector is not Lie dragged: $\pounds_l \eta^\m = l^\n\na_\n \eta^\m-\eta^\n \na_\n l^\m+2 C^\m{}_{\s\n} \eta^\s l^\n = 2 C^\m{}_{\s\n} \eta^\s l^\n $. In spite of the fact that in the presence of torsion differential parallelograms do not close, it is natural to still require the Lie dragging of $\eta^\m$ (see e.g. \cite{Luz:2017ldh}). With this definition of the congruence, shear and expansion are never modified by torsion, but only the twist. The NP spin coefficients $\os{\s}$ and $\os{\r}$ lose their geometric interpretation.}

\subsection{Torsionlessness of the affine null congruence} 

In this subsection we use the affine congruence defined above to study the geometric meaning of the various constraints present in the theory, in particular those responsible for the metricity of the congruence.
Let us begin with the Gauss constraint $\cal G$ in \Ref{primary}. First, we decompose it into rotations $L_i := \f12 \eps_{ijk} {\cal G}^{jk}$ and boosts $K_i := {\cal G}_{0i}$. Then, we consider the projections along $\chi^i$, and perpendicular to it, defined by $v_\perp^i:=\eps^{ijk}\chi_j v_k$ (notice that $v_\perp^i = -i v^-$). 
These various components have the following explicit forms (see Appendix~\ref{completegf}), 
\begin{subequations}\label{Gdecomp}\begin{align}
& L_\chi := \f12 \eps_{ijk} \chi^i {\cal G}^{jk} \overset{\vphi}\approx \eps_{ijk} \chi^i \tE^{aj} \eta_a^{k}, \\
& L_\perp^i := {\cal G}^{ij}\chi_j \overset{\vphi}\approx \p_a \tE^{ai} +\tE^{ai} \eta_a\chi - \eta_a^i \tE^a_j\chi^j - \tl\omega^i +\chi^i(\tl\om_j\chi^j-\p_a\tE^a_j\chi^j), \\
& K_\chi := \chi^i {\cal G}_{0i} \overset{\vphi}\approx \p_a\tE^a_i\chi^i - {\cal X}_{ij}\tE^{ai}\eta_a^j, \\
& K_\perp^i := \eps^{ijk} \chi_j {\cal G}_{0k} \overset{\vphi}\approx \eps^{ijk} \chi_j (\p_a\tE^a_k + \tE^a_k \,  \eta^i_a\chi_i -\tl\omega_k),
\end{align}\end{subequations}
where $\overset{\vphi}\approx$ means on-shell of the $\vphi$ constraint only, namely assuming $\chi^i$ constant. 

The two second class constraints are the linear combinations $\hat T_\perp^i := L_\perp^i-\eps^i{}_{jk}\chi^j K_\perp^k$, whose action would change the internal direction $\chi^i$. On the other hand, $T_\perp^i := L_\perp^i + \eps^i{}_{jk}\chi^j K_\perp^k$ and $L_\chi$ belong to the ISO(2) subgroup stabilising $x_+^I$ and are first class, together with $K_\chi$.\footnote{Covariantly, the stabilisers can be written as
 $T^I := 1/2 \eps^{IJKL} x_{+J} J_{KL}$ and $\hat T^I := -1/2 \eps^{IJKL} x_{-J} J_{KL}$. With $\chi^i=(1,0,0)$ and $M=2,3$, we have $v_\perp^M = (-v^3,v^2)$; the second class constraints read $\hat T^M=(L^2+K^3, L^3-K^2)$, whereas the first class ones are $T^M=(L^2-K^3, L^3+K^2)$.}
Using the explicit expressions for the spin coefficients (see Appendix \ref{AppSC}), we immediately identify the $x_+^I$-stabilisers with
\be\label{LTT}
L_\chi = 2 E \cN \im(\os{\r}), \qquad L_\perp^-+\eps^{-jk}\chi_jK_\perp^k = E \cN^2 \os{\k}\,.
\ee
These first class constraints are thus responsible for the congruence being geodesic and twist-free.
For the remaining first class constraint, we have
\be
K_\chi = E( E^a_i\chi^i \p_a \ln |E| - N\os{\th}_{(l)} +\p_a E^a_i\chi^i ).
\ee
Recalling that $\sqrt{-g} = NE$ and the explicit form of $l^\m$ from \Ref{coNP}, we see that this constraint puts the expansion on-shell:
\be
\os{\th}_{(l)} \approx \th_{(l)} = l^\m \p_\m \ln  \sqrt {-g} + \p_\m l^\m.
\ee
Let us remark the central role played by the radial boost: as a constraint, it is responsible for the metricity of the expansion; as a symmetry generator, it rescales lapse as discussed in \Ref{deltaN}.\footnote{As well as transforming the connection component determining lapse,
$\{K_\chi(\l), (\eta_a\chi)(E^a\chi)\} = \p_r \l$. At the level of covariant field equations, the radial boost constraint corresponds to the equation $\eps^{abc}\eps_{ijk} e^j_c\chi^k D_a e_b^i =0$.}
Finally, the second class constraints fix the two components of $\tl\om^i$ orthogonal to $\chi^i$, and have no direct implication for the affine congruence.

Let us now come to the non-affinity: even on-shell of the Gauss law, the now-geodesic congruence still carries non-affinity $k_{(l)}$, in spite of $l_\m$ being the gradient of a scalar. This is because the Gauss law only captures half of the torsion-less conditions. Where is then the equation setting $ k_{(l)}=0$? It must come from the Hamiltonian equation of motion that gives the stability of $\vphi^i\chi_i$, namely the equation capturing the fact that the level sets of $u$ provide a null foliation. Indeed, this stability condition was identified  in \cite{IoSergeyNull} as the multiplier equation expressing lapse in terms of canonical variables,\footnote{Recall that on a null hypersurface, the Hamiltonian constraint is second class, therefore its Lagrange multiplier satisfies an equation of motion, which fixes it up to zero modes. Concerning the zero modes, these are the left-over diffeomorphisms that on $\scri$ become the supertranslations of the BMS transformations.} which reads
\be\label{Eqlapse}
E^a_i\chi^i (\p_a\ln\cN - \eta_a^i\chi_i) =0. 
\ee
Comparing this expression with the first of \Ref{kk}, we see that it implies the vanishing of the non-affinity. 

It remains to put on shell the connection shear. To that end, we look at the light-cone conditions \Ref{Psi2}.
With our gauge-fixing $\chi^i=(1,0,0)$ the two components of $\hat\Psi^{ij}$ are $\hat\Psi^{23}$ are $\hat\Psi^{22}-\hat\Psi^{33}$. We combine them into a single complex equation, which gives
\be\label{Psilightcone}
-\f12 \Big(\Psi^{23} + \f i2(\Psi^{22}-\Psi^{33}) \Big) = \widetilde{N} \, \os{\s}_{(l)} 
-E^a_-E_b^-\p_a\tE^b_1 + E^a_1E^-_b\p_a\tE^b_- = 0,
\ee
from which it follows that
\be
\os{\s}_{(l)} = \f1N E^-_b (E^a_-\p_a E^b_1 - E^a_1\p_a E^b_-) = l^\m m^\n (\p_\m m_\n - \p_\n m_\m) = \f12 m^\m m^\m \pounds_l \g_{\m\n} \equiv \s_{(l)},
\ee
where in the second equality we used the explicit expressions \Ref{coNP} for the NP tetrad.
Hence, the two secondary simplicity constraints corresponding to the light-cone conditions make the connection shear metric.
Comparing this result with the analysis in metric variables of \cite{Torre:1985rw}, we expect the connection shear to be the conjugate momentum to the conformal metric. This expectation is indeed borne out, as we will show below in Section \ref{SecSympl}.

Summarising, 
the congruence generated by $l^\m$ is made geodesic by three first-class Gauss constraints. The fourth first-class one gives the relation between the connection expansion and the metric expansion. All these conditions are automatically preserved under evolution in $u$, since there are no secondary constraints arising from the stabilisation of the Gauss law. As for the connection shear, its relation to the metric shear is realised by the light-cone secondary simplicity constraints, and they are not automatically preserved. Tertiary constraints are required, to whose analysis we turn next.

\subsection{Tertiary constraints as the propagating equations}\label{SecTert}
Let us now discuss the tertiary constraints \Ref{tertcon}, whose presence is something quite unfamiliar within general relativity, and which is due to the combined use of a first-order formalism and a null foliation: each feature taken individually introduces a secondary layer of constraints in the Hamiltonian structure. Perhaps even more surprising is which of the field equations are described by these constraints: the propagating Einstein's equations, namely the dynamical equations describing the evolution (in retarded time $u$) of the shear away from the null hypersurface. In fact, it was shown in \cite{IoSergeyNull} that
\be\label{Ups2}
\Upsilon^{ab}
= -\f1{2\cN} \Pi^{ab}_{cd} \left[ 4g\, \eps_{efh} \, g^{0e}g^{cf} (\perp {\bG}^{\rm T})^{dh} + E^c_i \, ( \bB^{d i} +N^d \bB^{0i} ) \right],
\ee
where in the first term we recognise the propagating Einstein's equations,  and 
\be\label{Bianchi}
\bB^{\mu I}:= \eps^{\mu\nu\rho\sigma}e_{\nu J} F_{\rho\sigma}^{IJ}(\om)\equiv 0
\ee 
denotes the algebraic Bianchi identities.
This means that in the first-order formalism, the only time derivative present in the propagating equations \Ref{dynEqs} can be completely encoded in algebraic Bianchi equations. 

The equivalence \Ref{Ups2} may appear geometrically obscure, and it is furthermore not completely trivial to derive as a tensorial equation. On the other hand, it becomes transparent using the Newman-Penrose formalism, as we now show.
To that end, let us first identify the propagating equations in the Newman-Penrose formalism.
A straightforward calculation of the propagating equations gives
\be\nonumber
m^\m m^\n \perp_{\m\n}^{\r\s} G_{\r\s}(\om,e) = m^\m m^\n G_{\m\n}(\om,e) = m^\m m^\n R_{\m\n}(\om,e) = 
- R_{lmnm}(\om,e) - R_{nmlm}(\om,e) \approx 2R_{lmmn}(e),
\ee
where in the last equality $\approx$ means on-shell of the torsion-less condition.\footnote{These equations are not be confused with Sachs' optical equations $R_{lmlm}$ and $R_{lml\bar m}$, which relate Weyl and Ricci to the variation of shear and twist \emph{along} the null hypersurface, not away from it. 
}
Next, let us look at the tertiary constraints in its form \Ref{tertcon}, and project it in the same way on $S$:
\be
m_am_b\Upsilon^{ab} = \f12m_am_b E^{(a}_i\eps^{b)ef}\left( F^{0i}_{ef}- \chi_j F^{ij}_{ef}\right).
\ee
First, we have that
\be
F^{0i}_{ef}(\om,e)- \chi_j F^{ij}_{ef}(\om,e) = -\f 2\cN n^\m e^{i\n} R_{\m\n ef}(\om,e).
\ee
Then, to obtain the hypersurface Levi-Civita symbol, we observe that $n^\m$ is the only vector with a $u$-component, therefore we can write\footnote{With conventions $\eps^{0123}=1$, $e= -1/4! \eps_{IJKL} \eps^{\m\n\r\s} e_\m^I e_\n^J e_\r^K e_\s^L.$}
\be
\eps^{def} = - e 6 l^{[d}m^{e}\bar m^{f]}.
\ee
Finally, using the fact that $m_a E^a_i e^{i\n}=m^\n$, we have
\be
\f1E m_a m_b \Upsilon^{ab} = -\f1{e} n^\m m^\n m_d \eps^{def} R_{\m\n ef}(e,\om) = 2 n^\m m^\n l^\r m^\s  R_{\m\n \r\s}(e,\om)
= 2R_{nmlm}(e,\om), 
\ee
which coincides with (minus) the propagating equations on-shell of the torsion-less condition,
\be
m_a m_b \Upsilon^{ab} \approx - E m^\m m^\n G_{\m\n}(e).
\ee

It is also instructive to see the explicit role played by the algebraic Bianchi identity.
For vanishing torsion and NP gauge,\footnote{Namely $\r=\bar\r$, $\k=\eps=\pi=0$, $\t=\bar\a+\b$. See Appendix~\ref{AppNU} for details.} the propagating equation reads
\be
\Delta\s-\d\t +\bar\l\rho+(\m+\bar\g-3\g)\s +2\b\t + \Phi_{02} = 0,
\ee
where we can further set $\Phi_{02}=0$ since we are interested in the vacuum equations.
Here $\Delta:=n^\m\na_\m$ and $\d:=m^\m\na_\m$ is conventional NP notation, see Appendix \ref{AppSC}. For an expression of this equation in metric language, see e.g. \cite{Winicour16}.
The point is that if the connection is initially independent from the metric, this is a PDE with a single time derivative in the term $\D\s$; 
but this term can be eliminated using an algebraic Bianchi identity, or `eliminant relation' in the terminology of \cite{Chandra}. 
Using equation $(g)$ on page 48 of \cite{Chandra}, which in NP gauge reads
\be\label{Totti}
D\l + \D\bar\s-\bar\d(\a+\bar\b) = \bar\s(3\bar\g-\g+\m-\bar\mu) - 2\bar\b(\a+\bar\b),
\ee
we can replace $\D\s$ with $\d(\bar\a+\b)-D\bar\l$ plus squares of spin coefficients.
In metric variables, this would indeed be a trivial manipulation, since the time derivative is now simply shifted from $\D\s = - m^\m m^\n \p_u\p_r \g_{\m\n}+\ldots$ to $D\bar\l = - m^\m m^\n \p_r\p_u\g_{\m\n}+\ldots$. But used in the first order formalism with an independent connection (where now \Ref{Totti} holds with all $\os{\s}$ quantities and it is derived from \Ref{Bianchi}), relates non-trivially the propagating equations to the tertiary constraint.

Finally, concerning the geometric interpretation of this constraint, recall from Section \ref{Sec32} that it is there to stabilise the light-cone conditions: hence, Einstein's propagating equations can be seen as the condition that a metric-compatible connection shear on the initial null slice, remains metric at later retarded times.\footnote{This can be compared with the metric formalism of \cite{Torre:1985rw}, where the propagating Einstein's equations also arise from the stabilisation of the light-cone shear-metric conditions, but as multiplier equations, not as constraints.}

\section{Bondi gauge}\label{SecBg}

The discussion in the last two Sections has been completely general: apart from the condition of having a null foliation, we have not specified further the coordinate system. 
We now specialise to Bondi coordinates, presenting the simplified formulas that one obtains in this case.
We will then use this gauge to prove the equivalence of the symplectic potentials of the first-order and metric formalisms, which in particular identifies the connection shear with the momentum conjugated to the conformal 2d metric; and to discuss a property of radiative data at $\scri^+$. 

To that end, we completely fix the internal gauge, adapting the doubly-null tetrad to a $2+2$ foliation. For the interested reader, the Bondi gauge for our tetrad without the complete internal gauge-fixing is described in Appendix~\ref{AppNT}.
We take $\chi^i=(1,0,0)$ as in \Ref{chifixed}, and use the first-class generators $K_\chi$ and $T_M$ to fix $E^r_i = (1,0,0)$. This internal `radial gauge' adapts the tetrad to the $2+1$ foliation of constant-$r$ slices:
\be\label{intgf}
\chi^i=(1,0,0), \quad E^r_i = (1,0,0) \quad \Rightarrow \quad E_a^1=(1,0,0), \quad E = \sqrt{\g}, \quad m^\m=(0,0,E^A_-).
\ee
The determinant of the triad now coincides with that of the induced metric $\g_{AB}$ (hence triad and metric densities now conveniently coincide). This fixes five of the internal transformations, leaving us with the SO(2) freedom of rotations in the 2d plane of mappings $m^\m\mapsto e^{i\d}m^\m$. We will not use this freedom in the following, and if desired can be fixed for instance requiring the triad to be lower-triangular.
Now we impose the coordinate gauge-fixing.
On top of the null foliation condition $g^{00}=0$, the Bondi gauge conditions are $g^{0A}=0$, plus a condition on $r$, typically either the areal choice $\sqrt\g = r^2 f(\th,\phi)$, or the affine choice $g^{01}=-1$. We take here the affine Bondi gauge, and report the details on Sachs areal gauge in Appendix~\ref{AppSachs}. From the parametrisation \Ref{ginv}, we can read these conditions in terms of our tetrad variables:
\be\label{Bg}
g^{0a}=-\f1{\cN}E^a_i\chi^i = (-1,0,0). 
\ee
Using the internal gauge-fixing \Ref{intgf}, $E^r_i\chi^i =E^r_1=1$, hence \Ref{Bg} implies $E^A_1=0$ and $N=1$, as in the metric formalism.
The metric \Ref{g} and its inverse reduce to the following form,
\be\label{gU}
g_{\mu\nu}= \left(\begin{matrix} 2U + \g_{AB} N^AN^B & -1 & \g_{AB}N^B \\ & 0 & 0 \\
&& \g_{AB}\end{matrix}\right),
\qquad
g^{\mu\nu}= \left(\begin{matrix} 0 & -1 & 0 \\ & -2U & N^A \\
&& \g^{AB}\end{matrix}\right),
\ee
where we redefined $2U := -1 -2N^r$ for convenience. The triad and its inverse are
\be
E_a^i = \mat{1}{0}{-E^M_A E^A_1}{E^M_A}, \qquad E^a_i = \mat{1}{0}{E^A_1}{E^A_M},
\ee
where as before we use $M=2,3$ for the internal hypersurface coordinates orthogonal to $\chi^i$, and $E^A_M$ is the inverse of the dyad $E_A^M$.

The structure of the null congruence of $l_\m$ reduces to:
\begin{align}\label{LTTBg}
& k_{(l)} = -\eta_r^1, \qquad \os{\k} = \eta_r^-, \\
& \os{\s}_{(l)AB} = \eta_{(A}^M E_{B)M}, \qquad \os{\th}_{(l)} = \eta_A^M E_M^A, \qquad \os{\om}_{(l)AB} = \eta_{[A}^M E_{B]M}.
\end{align}
The lapse equation \Ref{Eqlapse} simplifies to
\be\label{EqlapseBg}
\eta_r^1 = 0, 
\ee
so this connection component is set to zero by working with a constant lapse.
The vanishing of the twist imposed by $L_\chi$ (in absence of torsion) now reads
\be\label{etaAS}
\eta_{[AB]} := \eta_{[A}^{M}\tE^{M}_{B]} = 0.
\ee
This equation is the null-hypersurface analogue of the familiar symmetry of the extrinsic curvature in the spatial hypersurface case, there analogously imposed by part of the Gauss constraint: $K_{[ab]}:=K^i_{[a} E^i_{b]}=0$.
The radial boost $K_\chi$ simplifies to
\be
K_\chi = \sqrt{\g} \Big(\os{\th}_{(l)} - \p_r\ln\sqrt{\g}\Big),
\ee
and its solutions give the affine Bondi-gauge formula for the expansion, $\os{\th}_{(l)}=\th_{(l)}:=\p_r\ln\sqrt{\g}$.
The solution of the light-cone secondary simplicity constraints \Ref{Psilightcone} now gives 
\be
\os{\s}_{(l)} = E_A^-\p_r E^A_-, 
\ee
namely the expression for the shear in affine Bondi gauge, written here in terms of the dyad $E^A_M$.

\subsection{Equivalence of symplectic potentials}\label{SecSympl}
We now show the equivalence between the symplectic potential in connection variables (which we can read from the $p\d q$ part of \Ref{actionH}) and the one in metric variables \Ref{Th1}, thereby identifying the canonical momentum to the conformal 2d metric in the connection language. It will turn out to be the connection shear of the canonical normal $n^1_\m=N l_\m$, as to be expected from the on-shell equivalence of the first and second order pure gravity action principles. 
As in the usual space-like canonical analysis, the equivalence of symplectic potentials will require the Gauss law. 
We begin by eliminating $\chi^i$ and $\pi^{ij}$ from the phase space, completely fixing the internal gauge and using the  primary simplicity constraints, and consider then only the first term of \Ref{actionH} for the symplectic potential.
Since our main focus are the bulk physical data, we will neglect boundary contributions to the symplectic potential, and show the equivalence in the partial Bondi gauge $g^{0A}=0$. The reason not to fix completely the Bondi gauge is to keep both lapse and an arbitrary $\sqrt{\g}$, to show a more general equivalence holding regardless of the choice of coordinate $r$. 
Hence, we want to show that
\be\label{Th=}
\Th = \int_\Si 2\tE^a_i\d\eta_a^i \overset{(g^{0A}=0)}{\approx} \int_\Si \sqrt{\g} \, \Pi^{AB}\d\g_{AB},
\ee
with $\Pi_{AB}$ given by \Ref{PiTorre}. 

The partial Bondi gauge is $E^A_i\chi^i=E^A_1=0$, which implies $\eta_r^M=0$ on shell of the Gauss law, see \Ref{LTTBg}. This eliminates two monomials from the integrand, and we are left with the following two terms:
\be
\Th = \int_\Si 2\tE^a_i\d\eta_a^i \approx \int_\Si 2(\tE^A_M\d\eta_A^M + \sqrt{\g}\d\eta_r^1).
\ee
Accordingly, here and in the following we will restrict attention to variations preserving the gauge and the Gauss constraint surface.
Let us look at the right-hand side of \Ref{Th=}.
We expect from the metric formalism that the conjugate momentum is build from the congruence of $n_1^\m = Nl^\m$. Its shear and expansion  are just $N$ times those of $l^\m$, which we can read from \Ref{lNGC}; its non-affinity is $k_{(n_1)}=\p_r\ln N = \eta_r^1$ using the lapse equation \Ref{Eqlapse} in partial Bondi gauge. Accordingly, we consider the following ansatz for the momentum,
\be\label{PiConn}
\Pi_{AB} := \eta_A^M E_{BM} - \g_{AB} (E^A_M \eta_A^M + \eta_r^1), 
\ee
whose decomposition gives
\be
\Pi_{AB} - \f12\g_{AB} \Pi = \eta_{(A}^M E_{B)M} - \f12\g_{AB}\os{\th}_{(n_1)} \equiv \os{\s}_{(n_1)AB}, \qquad \Pi = -\os{\th}_{(n_1)} - 2\eta_r^1,
\ee
where we used $\eta_r^M=0=\eta_{[A}^M E_{B]M}$ from the Gauss law.
This momentum reduces to the one in the metric formalism \Ref{PiTorre} by construction, and we now show it satisfies \Ref{Th=}.
To that end, we first observe that $E=\sqrt{\g}$ is now a $2\times 2$ determinant. This means that $\det E_A^M=\det \tE^A_M$,
and the inverse induced metric has the following expression in terms of canonical variables,
\be
\g^{AB} = \f{\tE^A_M\tE^{BM}}{(\det\tE^A_M)^2}.
\ee
A simple calculation then gives
\begin{align}\nonumber
\Pi^{AB}\d\g_{AB} &= - \Pi_{AB}\d\g^{AB} = -2\left[ \Pi_{(AB)}E^{BM} - \Pi E^M_A\right] \d\tE^A_M
\\ &= -2\left[ \eta_{(A}^M E_{B)M} E^{BN}\d\tE^A_N + \eta_r^1 E^M_A\d\tE^A_M \right].
\end{align}
where we used  $\d\det\tE = E^M_A\d\tE^A_M$.
Next, we use again $\eta^M_{[A}\tE_{B]M}=0$ from the Gauss law, so the first symmetrised term above gives twice the same contribution. Using the fact that $\d\sqrt{\g} = \cN E^M_A\d\tE^A_M$, we finally get
\be
\Pi^{AB}\d\g_{AB} \approx -2\left[ \eta_{A}^M \d\tE^A_M +\eta_r^1\d\sqrt{\g}\right],
\ee
and \Ref{Th=} follows up to boundary terms. We have thus verified that in the first order formalism the (traceless part of the) conjugate momentum to the induced metric is the connection shear of $n_1=Nl_\m$.

We also remark the presence of a term proportional to the 2d area. As in the metric formalism, this is a measure-zero degree of freedom, that can be pushed to a corner contribution and describes one of Sachs' corner data. A similar corner term appears in the spinorial construction of \cite{Wieland:2017cmf}, where it is shown to admit a quantisation compatible with that of the loop quantum gravity area operator. See also \cite{Freidel:2015gpa} for related results on 2d discreteness.

This result provides an answer to one of the open questions of \cite{IoSergeyNull}, namely that of identifying the Dirac brackets for the reduced phase space variables. We did so looking at the symplectic potential as in covariant phase space methods, and completely fixing the gauge: this introduced additional second class constraints that could be easily solved, e.g \Ref{EqlapseBg}. Whether it is possible to write covariant Dirac brackets without a complete gauge-fixing remains an open and difficult question, because of the non-trivial field equations satisfied by the second class Lagrange multipliers. 

It is interesting to compare the situation with the space-like case, where the dynamical part of the connection is also contained in components of $\eta^i_a$, except now $\chi^i$ belongs to a time-like 4-vector (and we can always set $\chi^i=0$, a choice often referred to as `time gauge', since $e^0\propto dt$). These dynamical components describe boosts and therefore do not form a group. An SU(2) group structure can be obtained via a canonical transformation, to either complex self-dual variables, as in the original formulation \cite{Ashtekar:1986yd}, or to the auxiliary Ashtekar-Barbero  real SU(2) connection (see e.g. \cite{ThiemannBook}): the transformation requires adding the Immirzi term to the action, and the price to pay is either additional reality conditions, or use of an auxiliary object instead of a proper spacetime connection. 
Using a null foliation appears to improve the situation: 
the three internal components of $\eta^i_a$ can be naively\footnote{To make the argument precise, we should embed the dynamical components into a covariant connection whose non-dynamical parts are put to zero by linear combinations of constraints, see e.g. \cite{AlexandrovLivine} for an analogue treatment in the space-like case.} associated with the radial boost $K_\chi$ and the two `translations' $T^i_\perp$, or null rotations, related to the ISO(2) group stabilising the null direction of the hypersurface. But as we have seen above only the translation components $\eta_a^M$ enter the bulk physical degrees of freedom, which are described by the connection shear. The component $\eta_a^i\chi_i$ is on a different footing: it enters the spin coefficients $\a,\b,\g$ and $\eps$ (see Appendix \ref{AppSC}), and is treated in a way similar to the expansion $\th$, in that it is fully determined from initial data on a corner.
We plan to develop these ideas in future research, in particular investigating the relation with a loop quantum gravity quantization based on the translation components of the connection, representing bulk physical degrees of freedom.

To complete the comparison between null and space-like foliations, in the latter case the 
canonical momentum conjugated to the induced metric is build from the triad projection $K^i_a$ of the extrinsic curvature (see e.g. \cite{ThiemannBook} for details). For a null foliation,  
the canonical momentum conjugated to the induced metric is related to the shear of the null congruence. The comparison is summarised by the following table:\footnote{For the reader interested in the time-like case, see \cite{AlexandrovKadar05}.}
\begin{spacing}{1.8}
\begin{center}\begin{tabular}{|l||l|l|}
\hline foliation & {\bf space-like} & {\bf null} \\ \hline
relevant internal group & SU(2) & ISO(2) \\
momentum conjugated to metric & $\Pi_{ab}=K_a^i E_{bi} - q_{ab} K_c^i E^c_i$ & $\Pi_{ab}={\cal X}_{ij} (\eta_a^i E_b^j - q_{ab} \eta_c^i E^{cj})$ \\
\hline
\end{tabular}\end{center}
\end{spacing}
To help the comparison in the table above, we have used the fact that in our formalism we can define the raised-indices hypersurface metric $q^{ab}$, and use it to prescribe an extension $\Pi_{ab}$ of \Ref{PiConn} on the whole hypersurface.\footnote{The equivalence \Ref{Th=} can then be written with $\Pi^{ab}\d q_{ab}$ on the right-hand side, and trivially holds because the extra pieces now present are put to zero by the constraints and/or gauge conditions.}

\subsection{Radiative data at future null infinity as shear `aligned' to $\scri^+$}\label{SecScri}
As a final consideration, we would like to come back to the geometric interpretation of the tertiary constraints, and point out that the very same algebraic Bianchi identity that links them to the propagating equations, also plays an interesting role in the interpretation of the radiative data at $\scri^+$. 

To that end, we consider in this subsection the case of an asymptotically flat spacetime, and the $u=$constant null foliation attached to future null infinity $\scri^+$. 
In this setting, we can compare our metric \Ref{gU} and doubly-null tetrad to those of Newman-Unti \cite{Newman:1962cia,NewmanTod,Adamo:2009vu} mostly used in the literature, and use the asymptotic fall-off conditions for the spin coefficients there computed.\footnote{In using these results, care should be taken in that the authors use a slightly different definition of coordinates: $u$ is now $1/\sqrt{2}$ the retarded time, and $r$ is $\sqrt{2}$ the radius of the asymptotically flat 2-sphere.} We refer the interested reader to Appendix~\ref{AppNT} for the details, and report here only the most relevant results. 
In particular, 
\be
\s = \f{\s^0}{r^2} + O(r^{-4}),
\ee
and the asymptotic shear $-\s^0(u,\th,\phi)$ fully characterises the radiative data at $\scri^+$ \cite{Penrose:1962ij, NP62,Ashtekar:1981hw}.
Ashtekar's result \cite{Ashtekar:1981hw} (see also \cite{Ashtekar:2014zsa} for a recent review) is that the data can be described in terms of a connection $D_\m$ defined intrinsically on $\scri^+$, related to the shear by
$\s^0_{\m\n}= D_\m l_\n - \f12 \g_{\m\n} \g^{\r\s}D_\r l_\s$. This description has led to a deeper understanding of the physics of future null infinity, showing among other things that the phase space at $\scri^+$ is an affine space (there is no super-translational invariant classical vacuum).
The connection description at $\scri^+$ inspired and is exactly analogous to the local spacetime connection description studied in this paper.

\begin{figure}[ht]   
\centering        \includegraphics[width=8cm]{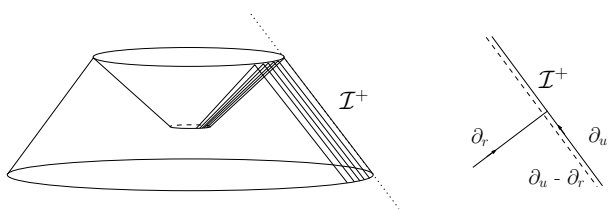}
\caption{\label{FigScri} {\small{
\emph{Characteristic initial-value problem at $\scri^+$. One prescribes data on a chosen $u_0$ hypersurface of the foliation attached to future null infinity, plus the asymptotic transverse shear $-s^0(u,\th,\phi)$. Thanks to the algebraic Bianchi identity \Ref{Totti}, this can also be understood as prescribing a certain shear for the non-geodetic asymptotic null vector 
$\p_u-\p_r$ in the physical spacetime.
} }} }
\end{figure}

From the perspective of the $2+2$ characteristic initial-value formulation (with backward evolution -- or we should rather say final-value formulation), this means that one can think of $\scri^+$ as one of the two null hypersurfaces, but the relevant datum there is not the shear along it (which vanishes!), but the transverse asymptotic shear $-\s^0(u,\th,\phi)$ at varying $u$, see Fig. \ref{FigScri}. However, we now show that thanks to the Bianchi identity \Ref{Totti}, this datum can also be identified as shear of a vector field in the physical spacetime. 

To that end, consider the second null vector of the tetrad, $n^\m$. It is null everywhere but non-geodesic, with
\be
n^\n\na_\n n^\m = 
-{\g}\, n^\m + {\n} \, m^\m + {\rm cc}.
\ee
In the asymptotic expansion, 
\be\label{nAs}
n^\m\p_\m \stackrel{r\mapsto\infty}{\longrightarrow}\p_u-\p_r
\ee
is leading-order twist-free and affine, but still non-geodesic:
\be
\om_{(n)} := \im(\m) = \bar m^\m m^\n \na_\n n_\m = O(r^{-2}), \qquad \g = O(r^{-2}), \qquad \n = \f{\psi_3^0}r + O(r^{-2}).
\ee
The non-geodesicity at leading order depends on one of the asymptotic complex projections of the Weyl tensor, in turn given by the radiative data $\psi_3^0 = \d \dot{\bar\s}^0$.\footnote{This can be seen solving at first order in $1/r$ the NP components $R_{n\bar m n l}$ and $R_{n\bar m n m}$ of the Riemann tensor, see e.g. (310i) and (310m) of \cite{Chandra}.}  
Since $n^\m$ is not geodesic, it is also not hypersurface orthogonal, in spite of being twist-free at lowest order: the radiative term $\d \dot{\bar\s}^0$ prevents the identification of a null hypersurface normal to \Ref{nAs} (except in the very special case of completely isotropic radiation at all times). Consequently, there is no unique definition of shear for the congruence it generates. Using the NP formalism, it is natural to consider the shear along the 2d space-like hypersurface spanned by $m^\m$, and define
\be
\s_{(n)} := -\l = - \bar m^\m \bar m^\n \na_\n n_\m = \f{\l^0}r + O(r^{-2}).
\ee

At the same lowest order $O(r^{-1})$, the algebraic Bianchi identity \Ref{Totti} can be solved to give 
\be\label{ls}
\l^0=\dot{\bar\s}^0,
\ee
which relates the transverse asymptotic shear to the $\l$-shear of $n^\m$. 
Hence, the radiative data at future null infinity correspond to a shear of a non-geodesic vector field `aligned' with $\scri^+$.
The fact that the vector is non-geodesic shows that the asymptotic $2+2$ problem can not be formulated in real spacetime.
On the other hand, this is how close one can get, in terms of the interpretation of the main constraint-free data, in bridging between the local 2+2 characteristic initial-value problem, and the asymptotic one.

\section{Conclusions}

In this paper we have presented and discussed many aspects of the canonical structure of general relativity in real connection variables on null hypersurfaces. We 
have clarified the geometric structure of the Hamiltonian analysis presented in \cite{IoSergeyNull}, explaining the role of the various constraints and their geometric effect on a null congruence. We have seen how the Lorentz transformations of the null tetrad are generated canonically, and how to restrict them so to adapt the tetrad to a $2+2$ foliation, and compare the connection Hamiltonian analysis to the metric one. Lack of canonical normalisation for a null vector means that the equivalence of the lapse functions can only be given up to a boost along the null direction.
Restricting to the Bondi gauge, we have identified constraint-free data in connection variables, and shown equivalence of the symplectic potential with the metric formalism. The metric canonical conjugated pair `conformal 2d metric/shear' is replaced in the first order formalism by a pair `densitized dyad/null rotation components of the connection', with the null rotations becoming the shear on-shell of the light-cone secondary simplicity constraints. 
In the presence of torsion, the connection can pick up additional terms that contribute to the shear, twist and expansion of the congruence, leading to modifications of Sachs' optical and Raychaudhuri's equations.

Even in the absence of torsion, the on-shell-ness is not automatically preserved under retarded time evolution, but requires of tertiary constraints, something unusual in canonical formulations of general relativity. We have shown that the tertiary constraints encode Sachs' propagating equations thanks to a specific algebraic Bianchi identity, the same one that allows one to switch the interpretation of the radiative data at $\scri^+$ from the transverse asymptotic shear $\s^0$ to the `shear' $\l^0$ of a non-geodetic, yet twist-free, null vector aligned with $\scri^+$, suggesting a different perspective on the asymptotic evolution problem.
The identification of the connection constraint-free data as null rotations means that the degrees of freedom form a group, albeit non-compact, hence one could try to use loop quantum gravity quantization techniques without introducing the Immirzi parameter. Some of the corner data, which we did not investigate here, have already be shown to lead to a quantization of the area \cite{IoNull,Wieland:2017zkf,Freidel:2015gpa}. A quantization of the connection description of the radiative degrees of freedom can lead to new insights both for loop quantum gravity and for asymptotic quantisations based on a Fock space.

We completed the paper with an extensive Appendix, presenting the explicit calculations of the first-order spin coefficients for the tetrad description used, and a detailed comparison between null tetrad descriptions and $2+2$ foliations.

We hope that the connection formalism can provide a new angle on some of the open questions on the dynamics of null hypersurfaces in general relativity, and we plan to come back in future research to some of the important aspects left open here: in particular, investigating the symplectic potential and Dirac brackets among physical data without the Bondi gauge, as well as including boundary terms and identifying the BMS generators in this Hamiltonian language. We also plan to develop further the indications that the connection degrees of freedom now form a group and its possible applications to quantisation.

\subsection*{Acknowledgments}
We are indebted to Sergei Alexandrov for many exchanges and a careful reading of the draft. We also would like to thank Abhay Ashtekar, Tommaso De Lorenzo, Michael Reisenberger and Wolfgang Wieland for helpful discussions.

\appendix
\section*{Appendix}
\setcounter{equation}{0}
\renewcommand{\theequation}{\Alph{section}.\arabic{equation}}

\section{Spin coefficients}\label{AppSC}
We use $\chi^i=(1,0,0)$ and $v^\pm:=(v^2\pm iv^3)/\sqrt{2}$.
For the tetrad derivatives we have
\begin{align}
& D = l^\m\na_\m= \f1\cN E^a_1 \na_a, \qquad \D = n^\m\na_\m=\f12 (\na_t -
(N^a+\f{\cN}2 E^a_1) \na_a),\\
& \d =m^\m\na_\m =E^a_- \na_a, \qquad \bar\d =m^\m\na_\m= E^a_+ \na_a\,.
\end{align}
For the spin coefficients we use the standard notation consistent with our mostly plus signature (which carries an opposite sign as to the notation with mostly minus signature) and use an apex $\circ$ to keep track of the fact that the  connection $\om^{IJ}_\m$ is off-shell. We then have
\begin{align}
& \os{\a} := -\f12 (n^\m \bar\d l_\m + m^\m \bar\d \bar m_\m)                                                                    
=\f12 E^a_+ \eta^1_a - \f i2 r_{1+} -\f14\om^+  -\f12 \bar\d\ln\cN \\
& \os{\b} := -\f12 (n^\m \d l_\m + m^\m \d \bar m_\m) 
= \f12 E^a_- \eta^1_a - \f i2 r_{1-} + \f14 \om^- -\f12 \d\ln\cN \\
& \os{\g} := - \f12 (n^\m \Delta l_\m + m^\m \Delta \bar m_\m)                  
= -\f14\cN E^a_1 \eta^1_a - \f12 N^a \eta^1_a +  \f i 4\cN r_{11}  +\f i2 N^a E_a^j r_{1j}+\f i4 N^a (E_a^l \eps_{1lm} \om^m)+ \\ \nn
&\qquad + \f12 (\om_0^{01}-i\om_0^{23}) -\f12 \Delta \ln\cN  \\
& \os{\eps} := -\f12 (n^\m D l_\m + m^\m D \bar m_\m)  
=\f1{2\cN} E^a_1 \eta^1_a - \f i{2\cN} r_{11} -\f12 D \ln\cN \\
& \os{\kappa} := -m^\m D l_\m =  -\f1{\cN^2} E^a_1 \eta^-_a\\
& \os{\tau} := -m^\m \Delta l_\m =  
\f12 E^a_1 \eta^-_a + \f{N^a}{\cN} \eta^-_a - \f{\sqrt 2}{\cN} (\om_0^{0-} - \om_0^{1-} ) \\
&\os{ \s} := -m^\m \d l_\m =  -\f1{\cN} E^a_- \eta^-_a \\
& \os{\r} := -m^\m \bar\d l_\m =   -\f1{\cN} E^a_+ \eta^-_a \\
\label{osm}
& \os{\m} := \bar m^\m \d n_\m   
= \f12 \cN \Big( E^a_- \eta^+_a - \om^1-i r_{22} -i r_{33} \Big) \\
&\os{\n } := \bar m^\m \Delta n_\m 
=-\f\cN4( \cN E^a_1 + 2 {N^a} )\eta^+_a -\f{({\cN})^2}2 (\om^+-ir_{1-})-\f{\cN}2N^a(E^1_a\om^+ - E^+_a\om^1 - 2iE^i_ar_{i+})+\\ \nn
&\qquad + \f1{2\sqrt{2}} \cN (\om_0^{0+}+\om_0^{1+}) \\
&\os{ \l} := \bar m^\m \bar\d n_\m 
=\f12 \cN \Big( E^a_+ \eta^+_a + 2 r_{23}-i r_{22} +i r_{33} \Big)  \\
&\os{ \pi} := \bar m^\m D n_\m  
 =\f12 E^a_1 \eta^+_a + \f12 \om^+ -i r_{1-} &
\end{align}
Under the rescaling $(l^\m,n^\m)\mapsto (l^\m/ A ,  A n^\m)$ (a class III transformation), 
\begin{align}
&\a \mapsto \a-\f1{2 A}\bar\d A, \qquad \b \mapsto \b-\f1{2 A}\d A, \qquad \g \mapsto  A\g-\f1{2 A}\D A, \qquad \eps \mapsto \f1 A\eps-\f1{2 A}D A, \\
&k \mapsto \f1{ A^2}k, \qquad \t \mapsto \t, \qquad \s \mapsto \f1 A\s, \qquad \r \mapsto \f1 A\r,\qquad \m \mapsto  A\m, \\
&  \n \mapsto  A^2\n, \qquad \l \mapsto  A\l, \qquad \pi \mapsto \pi,
\end{align}
Hence, many factors of $N$ disappear in the spin coefficients if we use the ADM-like normal $l^{\rm ADM}_\m = -N\p_\m u$.

\section{Congruence}\label{AppCongruence}
The complete expression of the congruence tensor with an affine connection is
\begin{align}\nonumber
& \nabla_0 l_0= \om^{0i}_0(\f1{\cN}{\cal X}_{ij}E^j_aN^a-\chi_i)+\f1\cN\om^{ij}_0\chi_j E_a^iN^a+\f1\cN\p_0\cN, \qquad
\na_a l_b = \f1\cN {\cal X}_{ij} \eta^i_a E^j_b, \\
& \na_0 l_a =\f1{\cN}(\om^{0j}_0{\cal X}_{ij} E^i_a +\om^{ij}_0\chi^j E^i_a), \qquad
\na_a l_0 = \eta_a^i(\f1\cN{\cal X}_{ij} N^aE_a^j - \chi^i) +\p_a\ln\cN
\end{align}
with projection $B_{\m\n} = \perp^\r{}_\m \perp^\s{}_\n \na_\r l_\s$ given by
\begin{align}
& B_{00}:=\f1N q^c{}_b N^aN^b \eta_{a}^M E^M_{c}, \qquad 
B_{0a}=\f1{N} q^b{}_{c} N^c\eta_{b}^M E^M_{a}, \qquad
B_{a0}=\f1{N} q^b{}_{a} N^c\eta_{b}^M E^M_{c}, \nn\\
& B_{ab}=\f1N q^c{}_a q^d{}_b {\cal X}_{ij}\eta_a^i E_b^j.
\end{align}

\section{Tetrad transformations and gauge fixings}\label{completegf}
At the Hamiltonian level, the Lorentz transformations are generated by the Gauss constraint ${\cal G}_{IJ}$, usually decomposed into spatial rotations
$L_i$ and boosts $K_i$, whose canonical form from \Ref{primary} reads
\begin{align}
L_i&:= \f12\eps_{ijk} {\cal G}^{jk} =\p_a (\eps_{ijk}\tE^a_j\chi^k)-{\eps_{ij}}^k\eta_a^{j}\tE^a_k-{\eps_{ij}}^k \tl\omega^j\chi_k,
\nonumber\\\label{GApp}
K_i&:= {\cal G}_{0i}
=\p_a\tE^a_i+ (\tE^a_i\chi_j-\tE^a_j\chi_i)\eta_a^{j}-{\cal X}_{ij}\tl\omega^j.
\end{align}
Since we are working on a null hypersurface, it is convenient to introduce the subgroups ISO(2) stabilising the null directions $x_{\pm}^I=(\pm 1,\chi^i)$, with generators 
 $T^I := 1/2 \eps^{IJKL} x_{+J} J_{KL}$ and $\hat T^I := -1/2 \eps^{IJKL} x_{-J} J_{KL}$. Both groups are 3-dimensional and contain the helicity generator $L_\chi$, plus two independent pairs of `translations',
$T_\perp^i:=\eps^{ijk}\chi_j T_k$ stabilising $x_+^I$, and $\hat T_\perp^i:=\eps^{ijk}\chi_j \hat T_k$  stabilising $x_-^I$.
Taking both sets and the radial boost $K_\chi$ we obtain the complete the Lorentz algebra, expressed in terms of canonical variables in \Ref{Gdecomp}.

For ease of notation and to make the formulas more transparent, we fix from now on $\chi^i=(1,0,0)$, as we did in most of the main text. We use the orthogonal internal indices $M=2,3$, and write the canonical form of the generators as follows,
 \begin{align}\nn
& L_1 = \eps_{1MN} \tE^{aM}\eta_a^N, && T_M=-\eps_{1Mi} \tE^a_1\eta_a^i 
 \\
& K_1 = \p_a\tE^a_1 - \tE^a_M \eta^M_a, && \hat T_M= -\eps_{1Mi} (\tE^a_1\eta_a^i -2\p_a\tE^a_i-2\tE^a_i\eta_a^1+2\tl\om^i).
 \end{align} 
To compute the action on the tetrad, we use the  brackets \Ref{PBe}. First of all, $\hat T_M$ change the internal null direction $\chi^i$:
\be
\{ \hat T_M, \chi_N\} = -\eps_{1MN}.
\ee 
Since the direction is gauge-fixed by \Ref{Cphi} in the action, these constraints are second class.

The stabilisers $T_M$ are first class, and can be used to put the triad in (partially) lower triangular form:
\be
\{ T_M, \tE^a_i \} = - \tfrac12 \eps_{1Mi} \tE^a_1,
\ee
so we can always reach $E^r_M=0$ with these transformations, and $E^1_A=0$ follows from the invertibility of the triad.
The radial boost $K_\chi$ can be used to fix $E^r_1=1$, since
\be
\{K_1, \tE^r_1\}=0, \qquad
\{ K_\chi, E\}=\tfrac12 E, \qquad
\{K_\chi, E^r_1\}=-\tfrac12 E^r_1.
\ee
The triad so gauge-fixed reads
\be\label{Elower}
E^i_a = \mat{1}{0}{E^M_r}{E^M_A}, \qquad E^a_i = \mat{1}{0}{E^A_1}{E^A_M}, 
\ee
where $E_A^M$ is the 2d dyad with inverse $E^A_M$, and $E^A_1=- E_M^A E_r^M$.
In this gauge, $d\phi^1=dr$, so the coordinates are adapted to the $2+2$ foliation. Furthermore, $E=\sqrt{\g}$ and so $\sqrt{-g}=NE=N\sqrt{\g}$.
Finally, the helicity rotation $L_1$, acting as 
\be
\{ L_1, \tE^a_i \} = \tfrac12 \eps_{1Mi} \tE^a_M, 
\ee
can be used to put to zero one off-diagonal component of the dyad and thus complete the triangular gauge of the triad.

Using hypersurface diffeomorphisms instead, we can put the triad in (partially) upper-triangular form:
\be
{\cal D}_a = 2\p_b (\eta_a^i\tE^b_i)-2\tE^b_i\p_a\eta_b^i+2\tl\omega^i\p_a\chi_i,\qquad
\{D(\vec N), \tE^a_i \} =  \pounds_{\vec N}\tE^a_i,
\ee
so we can use  ${\cal D}_A$ to fix $E^A\chi=0$, and ${\cal D}_r$ to fix $E^r_1=1$. This gives
\be\label{Eupper}
E^i_a = \mat{1}{E^1_A}{0}{E^M_A}, \qquad E^a_i = \mat{1}{E^r_M}{0}{E^A_M},
\ee
with $E^r_M = - E_M^A E^1_A$. In this gauge the hypersurface coordinates are not adapted to the $2+1$ foliation (the level sets $r=$constant do not span the 2d space-like surfaces), on the other hand the tangent to the null directions is now the coordinate vector $\p_r$.

For clarity, the various conditions that can be fixed using the various constraints are summarised in the table below, where by $r_{gf}$ we mean the final gauge fixing on $r$, for instance affine or areal.
\begin{center}
\begin{tabular}{|c|c|c|c|}
\hline $\cal H$ & ${\cal D}_A$ & ${\cal D}_r$ & \\
$g^{00}=0$ & $E^A\chi=0 \ \Leftrightarrow \ E_r^M=0$ & $E^r\chi=1$ aut $r_{gf}$ & \\ \hline 
$\hat T^i_\perp$ & $T^i_\perp$ & $K_\chi$ & $L_\chi$ \\ 
$\chi^i=(1,0,0)$ & $E_M^r=0 \ \Leftrightarrow \ E_A\chi=0$ & $r_{gf}$ aut $E^r\chi=1$ & $\d$ \\
\hline 
\end{tabular}
\end{center}

Notice that if one does not fix the upper or lower triangular form of the triad, the inverse of the 2d dyad if of course not given by the corresponding entries of the inverse triad. A general parametrisation of the triad in terms of the dyad can be easily written as follows,
\be\label{EApp}
E^i_a = \left(\begin{array}{cc} \hat M & {\cal E}^M_A f_M \\ {\cal E}^M_A\g^A & {\cal E}^M_A \end{array}\right),
\qquad E^a_i = \f1{M}\left(\begin{array}{cc} 1 & -f_M \\ -\g^A  & M{\cal E}^A_M+\g^A f_M \end{array}\right).
\ee
Here ${\cal E}_A^M$ is the dyad and ${\cal E}^A_M$ its inverse, $E = {\cal E} M$ and $M=\hat M - \g^A{\cal E}^M_A f_M$ is a $2+1$ lapse function.
Then $\g_{AB}={\cal E}^M_A{\cal E}_{MB}$ and
\be
q_{ab}=\mat{\g_{AB}\g^A\g^B}{\g_{AB}\g^B}{\g_{BA}\g^A}{\g_{AB}}.
\ee
The Bondi gauge sets $\g^A=0$, namely $q_{ra}=0$.

\section{$2+2$ foliations and NP tetrads}\label{Sec2+2}
We collect here various useful formulas relating the tetrad formalism to the $2+2$ foliation of \cite{dInverno:1980kaa} and \cite{Torre:1985rw}. 
As briefly explained in Section~\ref{SecH}, the $2+2$ foliation is induced by two closed 1-forms, $n^\a:=d\phi^\a$ locally, $\a=0,1$. These define a `lapse matrix' $N_{\a\b}$, as the inverse of 
$N^{\a\b}:=n_\m^\a n^{\b\m}$,
and a dual basis of vectors $n^\m_\a:=N_{\a\b} g^{\m\n} n^\b_\n$. Note that $n_0^\m$ and $n_1^\m$ are tangent respectively to the hypersurfaces 
$\phi^1=const$ and $\phi^0=const.$ We assume $\det N_{\a\b}<0$, so that the codimension-2 leaves $\{S\}$ are space-like. The projector on $\{S\}$ is $\perp^\m{}_{\n}:=\d^\m_\n - N_{\a\b}n^{\a\m}n^\b_\n$, and the covariant induced metric $\g_{\m\n}:=\perp_{\m\n}$. 
The 2d spaces $\{T\}$ tangent to $n_\a^\m$ are not integrable in generic spacetimes, since $ \perp^\m{}_\n[n_0,n_1]^\n\neq 0$. 
This non-integrability is often  referred to as twist in the literature.
On the other hand, the orthogonal 2d spaces foliate spacetime by construction, and we can introduce shift vectors to relate the tangent vectors to coordinate vectors, $b_\a^\m=(\p_{\phi^\a})^\m-n_\a^\m$.

To write the metric explicitly, we take coordinates $(\phi^\a,\s^A)$ adapted to the foliation, then
\be
g_{\m\n} = \mat{N_{\a\b}+\g_{AB}b_\a^A b_\b^B}{\g_{BC}b_\a^C}{\g_{AC}b_\b^C}{\g_{AB}}, \qquad 
g^{\m\n} = \mat{N^{\a\b}}{-N^{\a\b}b_\b^B}{-N^{\b\a}b_\a^A}{\g^{AB}+N^{\a\b}b_\a^A b_\b^B}.
\ee
For a null foliation, we fix one diffeomorphism requiring $N_{11}=0=N^{00}=g^{00}$, so that the first normal is null, and $N^{01}=1/N_{01}$, $N^{11}=-N_{00}/N_{01}^2$.
The norm of $n^1$ is $N^{11}$ and we leave it free (it can be both time-like or space-like without changing the fact that the orthogonal spaces $\{S\}$ are space-like), but notice that we can always switch to a null frame $(n^0,\tl n^1)$ with 
\be
\tl n^1 = N_{01} n^1+\f12 {N_{00}} n^0, \qquad |\!|\tl n^1|\!|^2=0, \qquad n^0_\m \tl n^{1\m} =1.
\ee
This can be used to define the first two vectors of a NP tetrad adapted to the foliation, via $l_\m:=-n^0_\m$, $n_\m:=\tl n^1_\m$, so that the 2d space-like induced metrics coincide
\be\label{folequi}
\g_{\m\n} = g_{\m\n} - N_{\a\b}n^{\a}_{\m}n^\b_\n = g_{\m\n} +2l_{(\m}n_{\n)}.
\ee
Notice that acting with a Lorentz transformation preserving $l$, we have
\be\label{classI}
n^\m\mapsto n^\m+\bar a m^\m +a\bar m^\m+|a|^2 l^\m,\qquad m^\m\mapsto m^\m+a l^\m;
\ee
one thus obtains a new covariant 2d metric, still space-like and transverse to $l_\m$, but not associated with the $2+2$ foliation any longer.
In terms of the NP tetrad, the non-integrability of the time-like spaces is measured by the two spin coefficients $\tau$ and $\pi$,
\be
m_\m[l,n]^\m=\t+\bar\pi.
\ee

\subsection{Adapting a NP tetrad}
We can also reverse the procedure: start from an arbitrary NP tetrad, and adapt it to a $2+2$ foliation. To that end, recall first that 
\be\label{mmbarInt}
[m,\bar m]^\n =(\m-\bar\m) l^\m+(\r-\bar\r)n^\m-(\a-\bar\b) m^\m+(\bar\a-\b)\bar m^\m,
\ee
so the general non-integrability of $(m,\bar m)$ is given by non-vanishing $\im(\r)$ and $\im(\m)$.
To adapt the NP to the $3+1$ null foliation, we choose $l:=-d\phi^0$. This fixes 3 Lorentz transformation, and implies $\im(\r)=0=\k$ and $\t=\bar\a+\b$.
We can also fix the SO(2) helicity rotation requiring $\eps=\bar\eps$. This leaves us with two tetrad transformations left. To have a $2+2$ foliation induced by the tetrad, we need  
\be\label{mu0}
\m-\bar\m=2m^\m \bar m^\n\p_{[\n}n_{\m]}=0.
\ee
This is achieved if in coordinates $(\phi^\a,\s^A)$ adapted to the foliation $m^\m=(0,0,m^A)$, hence $n_{\m}=(c_\a,0,0)$ by orthogonality; this fixes the remaining two tetrad freedoms (And if we fix radial diffeomorphisms to have $N^{01}=-1$, this gauge also implies $\pi=\a+\bar\b$).   
Inverting this linear system we find
\be
d\phi^0=-l, \qquad d\phi^1=\f{c_0}{c_1} l +\f1{c_1} n.
\ee
This identifies $c_\a=(N_{00}/2,N_{01})$, and \Ref{folequi} follows again. 
For more on the characteristic initial value problem in NP formalism see e.g. \cite{Racz:2013hva}. The use of a tetrad  adapted to a $2+2$ foliation is common, e.g. \cite{Sachs62,Hawking:1968qt,Winicour16}, but not universal. In particular in \cite{Sachs62} the partial Bondi gauge is completed with $N^{11}=0=N_{00}=c_0$, so to have both 1-forms $d\phi^\a$ null.

\subsection{The Bondi gauge and Newman-Unti tetrad}\label{AppNU}
A more wide-spread tetrad description, particularly suited to study asymptotic radiation, is the one introduced by Newman and Unti \cite{Newman:1962cia}, see e.g. \cite{NewmanTod,Adamo:2009vu} for reviews, which is adapted to the $3+1$ null foliation and to the Bondi gauge. We take coordinates $(u,r,\th,\phi)$ and fix $g^{00}=0$, so that the level sets of $u$ give a null foliation with normal $l_\m=-\p_\m u$. Recall that the null hypersurfaces $\Si$ normal to $l_\m$ are ruled by null geodesics, with tangent vector
\be
l^\m\p_\m = -g^{0\m}\p_\m = \f1N\p_r-g^{0A}\p_A.
\ee
This suggests a natural $2+1$ foliation of $\Si$ given by the level sets of a parameter along the null geodesics (affine or not). The description simplifies greatly if we gauge-fix $g^{0A}=0$, as to identify the geodesic parameter with the coordinate $r$, while simultaneously putting to zero the shift vector of the $r=const.$ foliation on $\Si$. In other words, the (partial) Bondi gauge $g^{0A}=0$ gives a physical meaning to the coordinate foliation defined by $u$ and $r$ by identifying it with the foliation defined by the null geodesics on $\Si$. In the $2+2$ language of \cite{dInverno:1980kaa,Torre:1985rw}, with adapted coordinate $\phi^0=u$, the gauge corresponds to a vanishing shift vector $b_1^\m$, so that $\p_{\phi^1}$ is tangent to the null geodesics.

Let us complete the Bondi gauge choosing affine parametrization, namely $g^{01}=-1$. The metric and its inverse read
\begin{align}
& g^{\m\nu} = \left(\begin{matrix} 0 & -1 & 0 \\ & g^{11} &  g^{1A} \\ && g^{AB} \end{matrix}\right),
\qquad g_{\m\nu} = \left(\begin{matrix}  -g^{11}+g_{AB} g^{1A} g^{1B} & -1 & g_{AB}g^{1B} \\ & 0 & 0 \\ && g_{AB}  \end{matrix}\right).
\end{align}

The Newman-Unti tetrad adapted to these coordinates is chosen identifying $l_\m$ with the normal to the foliation, and requiring $n^\m$ and $m^\m$ to be parallel propagated along $l^\m$.
It is parametrised as follows,
\be\label{eNT}
l^\m\p_\m = \p_r, \qquad n^\m\p_\m = \p_u+U \p_r + X^A\p_A, \qquad m^\m\p_\m = \om\p_r +\xi^A\p_A,
\ee
with $A=\z,\bar\z$ stereographic coordinates for $S^2$ ($\z=\cot\th/2e^{i\phi}$), and
\begin{align}
& g^{11}=2(|\om|^2-U),  
\qquad g^{1A}=\om \bar\xi^A+\bar\om\xi^A-X^A, 
\qquad g^{AB}= \xi^A\bar\xi^B+\bar\xi^A\xi^B.
\end{align}
The co-tetrad is
\be\nn
l_\m=(-1,0,0,0),\quad n_\m=\Big(U-g_{AB}X^A(\om\bar\xi^B+\bar\om\xi^B), -1, g_{AB}(\om\bar\xi^B+\bar\om\xi^B)\Big),
\quad m_\m=(-g_{AB}\xi^AX^B,0,g_{AB}\xi^B).
\ee

The coefficients are a priori 9 real functions $(U\in\R, \, X^A\in\R^2, \, \om\in\C, \, \xi^A\in\C^2)$ parametrising the 6 independent components of the metric plus 3 internal components corresponding to the ISO(2) stabiliser of $l^\m$. The helicity subgroup generates dyad rotations $\xi^A\mapsto e^{i\d}\xi^A$, and the translations the class $I$ transformations \Ref{classI}. The latter in particular shift $\om\mapsto\om+a$, $a\in\C$, and can be used to put $\om=0$, so $m^\m=(0,0,m^A)$ with 2d  space-like components only. This is the $2+2$-adapted choice described above, and corresponds to $E^r_M=0$ as in the lower-triangular form \Ref{Elower}, that we also used in Section~\ref{SecBg} in the main text to make easier contact with the metric Hamiltonian formalism.
Alternatively, this null rotation can be used to achieve $\pi=0$, so to make $n^\m$ and $m^\m$ to be parallel propagated along $l^\m$ as demanded by Newman and Unti.

In terms of spin coefficients, we have the following simplifications: $\k=\im(\r)=0,$ $\t=\bar\a+\b$ which follow from $l_\m$ being a gradient, $\re(\eps)=0$ from fixing the radial diffeos requiring $r$ affine parametrization, and $\pi=0$ from the parallel transport of $n^\m$ and $m^\m$.
Finally $\im(\eps)=0$ if we fix the helicity SO(2) rotation. This complete fixing is usually referred to as NP gauge, to be contrasted with the $2+2$-adapted gauge described above, where the condition $\pi=0$ is replaced by $\pi=\bar\t$ and $\im(\m)=0$. 

Hence, when we refer to the Newman-Unti tetrad \Ref{eNT} in NP gauge there are only 6 free functions of all 4 coordinates.
The NP gauge is preserved by class $I$ and helicity transformations with $r$-independent parameters.

\section{Mappings to the $\chi$-tetrad}\label{Appchi}
In this Appendix we discuss the detailed relation between the $\chi$-tetrad used to perform the canonical analysis in  real connection variables and the results of the previous Appendix. It provides formulas completing the discussion in the main text.

At the end of Section \Ref{Sec31} we introduced the internal `radial gauge' \Ref{nullgauge}, stating that it adapts the tetrad to the $2+2$ foliation and identifies the lapse function with the one used in the metric formalism. We now provide the relevant details and proofs.
The $\chi$-tetrad and its inverse are given by
\be\label{eApp}
e^I_\m = \left(\begin{array}{cc} \hat N & E^i_a\chi_i \\ N^a E_a^i & E^i_a \end{array}\right),
\qquad e^\m_I = \f1{\cN}\left(\begin{array}{cc} 1 & -\chi_i \\ -N^a  & {N} E^a_i+N^a\chi_i \end{array}\right),
\ee
where $\chi^2=1$ to have a null foliation, $e = E N$ and $N=\hat N -N^a E_a^i\chi_i$ is the lapse function.
Taking the soldered internal null directions $x_{\pm\m} = e^I_\m x_{\pm I}$ of \Ref{xpm}, and defining $m^\m$ to be a complex linear combination of the two orthogonal tetrad directions ${\cal X}^{ij}e_j^\m$, e.g. $m^\m:=\f1{\sqrt{2}}(e^\m_2-ie^\m_3)$ when $\chi^i=(1,0,0)$, the basis
\[
(x_{+}^\m,-x_-^\m,m^\m,\bar m^\m)
\] 
is a doubly-null tetrad. We then rescale it by
\be
l^\m = \f1\cN x_+^\m, \qquad n^\m=-\f\cN2 x_-^\m,
\ee 
to define an NP tetrad adapted to the $3+1$ null foliation as described in the main text, see \Ref{NP}. 
In general, the 2d spaces with tangent vectors $(m^\m,\bar m^\m)$ will not be integrable. With reference to \Ref{mmbarInt}, we see that integrability requires $\im(\r)=\im(\m)=0$. The first condition  is guaranteed by the fact that $l_\m$ is a gradient. The second can be obtained with a class $I$ transformation, generated by the translations ${\cal X}_{ij}T^j$ stabilising $l^\m$, fixing
\be\label{nullgaugeApp}
E_A\chi=0  \quad \Leftrightarrow \quad {\cal X}^{ij}E^r_j=0.
\ee
In this gauge 
\be
m^\m=(0,0,E^A_-),\qquad n_\m=(\cN(\cN/2+N^rE_r\chi), \cN E_r\chi,0,0),
\ee
so that the null tetrad is manifestly adapted to the $2+2$ foliation defined by the level sets of the coordinates $u$ and $r$.
Then $\im(\m)=0$ also follows immediately by explicit calculation of \Ref{mu0} using the fact that $m^\m$ only has 2d surface components. 

In a first-order formalism with independent connection, the statement holds in the absence of torsion. We have already seen in Section~\ref{SecNC} that on-shell of the torsionless condition $\im(\os{\r})=\im(\r)$. Let us show here explicitly how $\im(\os{\m})$ goes on-shell.
From \Ref{osm} we have 
\be
\im(\os{\m})=-\f{N^2}2\im(\os{\r})-\f N2(r_{22}+r_{33}),
\ee
and from one of the secondary simplicity constraints \Ref{noncovPsi} we have
\be
\Psi^{11} = -r_{22}-r_{33} - \eps^{1MN}\tE^a_M \Et^1_b\p_a\tE^b_N.
\ee
The last term vanishes for $E^r_M=0=E^1_A$, hence $\im(\os{\m})=0$ in this gauge.

To complete the comparison with the $2+2$ formalism, let us fix the internal direction $\chi^i=(1,0,0)$, and use $M=2,3$ to refer to the orthogonal directions. Then \Ref{nullgaugeApp} puts the triad in the form
\be\label{Eradial}
E^i_a = \mat{E^1_r}{0}{E^M_r}{E^M_A}, \qquad E^a_i = \mat{E^r_1}{0}{E^A_1}{E^A_M}, 
\ee
thus $E_A^M$ is the 2d dyad and $E^A_M$ its inverse, and we further have the equalities $E^1_r=1/E^r_1$, \linebreak $E_r^M=-E^M_AE^A_1/E^r_1$.
We then have $g_{AB}=q_{AB}=E^M_AE^{BM}=\g_{AB}={\cal E}_A^M{\cal E}_{BM}$, consistently with the fact that the metric induced by the dyad is adapted to the coordinates by the gauge-fixing, and $q^{AB}=E^A_ME^{BM}=\g^{AB}$ is its inverse. Notice that the (partial) Bondi gauge $g^{0A}=0$ achieves $g^{AB}=\g^{AB}$, analogously to the vanishing-shift gauge for space-like foliations. 

At this point $E=E_r\chi\sqrt{\g}$ and $\sqrt{-g}=E_r\chi N\sqrt{\g}$. A look at the metric shows that
\be
-1/g^{01} = E_r\chi N,
\ee
hence, the lapse function in the metric Hamiltonian analysis of \cite{Torre:1985rw}  equals the one in the connection formulation up to a factor $E_r\chi$. This ambiguity is not surprising due to the null nature of the foliation and the lack of a canonical normalization of its normal. To identify our lapse with the one in the metric formalism is sufficient to fix the radial boost $K_\chi$ as to have $E^r\chi=1$, as we did with \Ref{nullgauge}. Then also $E_r\chi=1$ because of \Ref{nullgaugeApp} and the triad takes the form \Ref{Elower}. We also recover the relation $\sqrt{-g}=N\sqrt{\g}$ between lapse and the determinant of the metric Hamiltonian analysis.
For completeness, we report below the relation between the $\chi$-tetrad coefficients and the $2+2$ foliation with a general radial gauge. The case with coinciding lapse functions can immediately be read plugging $E^r\chi=1=E_r\chi$ in the formulas below.

The relation between the foliating normals and the adapted null co-frame is given by 
\be\label{ni}
n^0 = du = - \f1{\cN} x_{+I} e^I, \qquad n^1 = dr = \f1{2E_r\chi}\Big( \f{\cN+2N^rE_r\chi}{\cN}x_{+I} + x_{-I} \Big)e^I, 
\qquad n^0_\m n^{1\m} = -\f1{E_r\chi\cN}.
\ee
The dual basis, shift vectors and lapse matrix are
\begin{align}
& n_0^\m = N(\tfrac{N^r}{E_r\chi}+\tfrac N2)l^\m+n^\m= (1, 0, N^rE_r\chi E^A\chi- N^A), && b_0^A = N^A-N^rE_r\chi E^A\chi,\\
& n_1^\m = E_r\chi N l^\m = (0, 1, E_r\chi E^A\chi), && b_1^A=-E_r\chi E^A\chi,
\end{align}
\be
N_{\a\b} = \mat{-\cN(\cN+2N^rE_r\chi)}{-\cN E_r\chi}{-\cN E_r\chi}{0},\qquad 
N^{\a\b} = \mat{0}{-\f1{\cN  E_r\chi}}{-\f1{\cN E_r\chi}}{\f1{\cN  (E_r\chi)^2}(\cN+2N^r  E_r\chi)},
\ee
and the formulas for the 2d projector and covariant induced metric coincide, 
\be\label{gindEqApp}
\g_{\m\n} := g_{\m\n} - x_{+(\m} x_{-\n)} = g_{\m\n} - N_{\a\b} n^\a_{\m} n^\b_{\n} = \mat{q_{ab} N^a N^b}{q_{bc}N^c}{q_{ab}N^b}{q_{ab}}.
\ee
From \Ref{Elower}, we see also that $E^A_1=-E^A_M E^M_r$, which provides an alternative characterisation of the second shift vector in terms of $E^M_r$.

The non-integrability of the $\{T\}$ surfaces is the same as measured by the null dyad, 
\be
\perp^\m{}_\n[n_0,n_1]^\n\equiv [n_0,n_1]^\m, \qquad m_\m [n_0,n_1]^\m = -N(\tau+\bar\pi).
\ee
Having gauge-fixed $N^{00}=0$ to have $du$ null and $r$ affine or areal, we cannot for general metrics simultaneously take $dr$ to be null. It can be made null on a single hypersurface $\tl\Si$ defined by some fixed value of $r=r_0$,  if we exploit the left-over freedom of hypersurface diffeomorphisms to fix $N^{11}=0$. This is what was done by Sachs in setting up the $2+2$ characteristic initial value problem, further fixing $N^A=0$ on the same hypersurface, so that the normal vector of $\tl\Si$ at $r=r_0$ is just $n_0^\m=n^\m=\p_u$, as in Fig.\ref{Fig}.

\subsection{The Bondi gauge and Newman-Unti tetrad}\label{AppNP}\label{AppNT}
In Section~\ref{SecBg} in the main text we discussed the Bondi gauge with a null tetrad already adapted to the $2+2$ foliation. This was motivated by the goal of recovering properties of the metric symplectic formalism. On the other hand, the Newman-Unti tetrad \Ref{eNT} mostly used in the literature is adapted to the $3+1$ null foliation only. In this Appendix we present the relation between our metric coefficients and those of \Ref{eNT} without fixing the internal `radial gauge' \Ref{intgf}.
To that end, we first fix all diffeomorphisms requiring the Bondi gauge
\be\label{BgApp}
\f1{\cN}E^a\chi =(1,0,0).
\ee
We then fix the internal direction $\chi^i=(1,0,0)$, and adapt $l=-du=x_+/N$. This leaves the freedom of acting with the ISO(2) subgroup stabilizing the direction. Because we rescaled the canonical tetrad by $N$, we also gain the freedom of canonical transformations corresponding to the radial boost $K_\chi$, which does not affect $l$. This additional gauge freedom should be fixed requiring $E^r\chi=1$, implying $N=1$.
We are then left with 9 free functions, 6 for the metric and 3 for the internal ISO(2) stabilising $l$. Comparing our tetrad \Ref{NP} in this gauge with \Ref{eNT} we immediately identify
\be
U = - \f12 -N^r, \qquad X^A = - N^A, \qquad \om = E^r_-, \qquad \xi^A = E^A_-.
\ee
The $2+2$-adapted tetrad is recovered with a class $I$ transformation setting $\om=E^M_-=0$.

\subsection{Areal $r$ and Sachs' metric coefficients}\label{AppSachs}
Above we used affine $r$, as usual in literature using the Newman-Penrose formalism. The alternative common choice is Sachs', leaving $g_{01}=-e^{2\b}$ free and requiring instead $\sqrt{\g}=r^2 f(\th,\phi)$.
Again we fix the internal direction $\chi^i=(1,0,0)$ and the radial boosts with $E^r\chi=1$, so to have the identification of our $N>0$ with the metric lapse $e^{2\b}$. 
The triad has the form \Ref{Eupper}, and the metric reads
\begin{align}
& g_{\mu\nu}= \left(\begin{matrix} -\cN(\cN+2N^r+2N^AE_A\chi)+\g_{AB}N^A N^B & -\cN & \g_{AB}N^B-\cN E_A\chi \\ & 0 & 0 \\
&& \g_{AB}\end{matrix}\right).
\end{align}
Comparing with \Ref{gS} in the main text, we find
\be
\b=\f12\ln{N}, \qquad U^A=-N^A+N\g^{AB}E_B\chi, \qquad \f Vr = 2N^1+ N(1 + \g^{AB}E_A\chi E_B\chi).
\ee
Reverting to affine $r$, $N=1$ and the map from Sachs' metric coefficients to Newman-Unti's is \linebreak $V/r = 2(|\om|^2-U)$, $U^A=X^A-\om\bar\xi^A-\bar\om\xi^A.$

\providecommand{\href}[2]{#2}\begingroup\raggedright\endgroup

\end{document}